\documentstyle[12pt,graphics,axodraw]{article}
\advance\textheight by \topskip
\begin{document}
\tolerance=100000
\thispagestyle{empty}
\setcounter{page}{0}

%       This defines et al., i.e., e.g., cf., etc.
\def\ie{\hbox{\it i.e.}{}}      \def\etc{\hbox{\it etc.}{}}
\def\eg{\hbox{\it e.g.}{}}      \def\cf{\hbox{\it cf.}{}}
\def\etal{\hbox{\it et al.}}    \def\vs{\hbox{\it vs.}{}}
\def\dash{\hbox{---}}
\newcommand\refq[1]{$\cite{#1}$}
\def\lq{{\rm LQ}}
\def\fb{~{\rm fb}}
\def\pb{~{\rm pb}}
\def\ev{\,{\rm eV}}
\def\mev{\,{\rm MeV}}
\def\gev{\,{\rm GeV}}
\def\tev{\,{\rm TeV}}
\def\wh{\widehat}
\def\wt{\widetilde}
\def\abs#1{\left| #1\right|}   
\newcommand{\slsh}{\rlap{$\;\!\!\not$}}     % Feynman slash
\newcommand{\as}{{\ifmmode \alpha_S \else $\alpha_S$ \fi}}
\newcommand{\ep}{\epsilon}
\newcommand{\be}{\begin{equation}}
\newcommand{\ee}{\end{equation}}
\newcommand{\bea}{\begin{eqnarray}}
\newcommand{\eea}{\end{eqnarray}}
\newcommand{\ba}{\begin{array}}
\newcommand{\ea}{\end{array}}
\newcommand{\bi}{\begin{itemize}}
\newcommand{\ei}{\end{itemize}}
\newcommand{\bn}{\begin{enumerate}}
\newcommand{\en}{\end{enumerate}}
\newcommand{\bc}{\begin{center}}
\newcommand{\ec}{\end{center}}
\newcommand{\ul}{\underline}
\newcommand{\ol}{\overline}
\newcommand{\sm}{${\cal {SM}}$}
\newcommand{\mssm}{${\cal {MSSM}}$}
\newcommand{\Dir}{\kern -6.4pt\Big{/}}
\newcommand{\Dirin}{\kern -10.4pt\Big{/}\kern 4.4pt}
\newcommand{\DDir}{\kern -7.6pt\Big{/}}
\newcommand{\DGir}{\kern -6.0pt\Big{/}}
\def\Ord{\buildrel{\scriptscriptstyle <}\over{\scriptscriptstyle\sim}}
\def\OOrd{\buildrel{\scriptscriptstyle >}\over{\scriptscriptstyle\sim}}
\newcommand{\lspage}[1]%   Landscape-page
{ 
\rotatebox{90}{ \begin{minipage}{\textheight} #1 \end{minipage} } 
  }
\def\epem{\ifmmode e^+e^-\else $e^+e^-$ \fi}
\relax
\def\ap#1#2#3{
        {\it Ann. Phys. (NY) }{\bf #1} (19#3) #2}
\def\app#1#2#3{
        {\it Acta Phys. Pol. }{\bf #1} (19#3) #2}
\def\cmp#1#2#3{
        {\it Commun. Math. Phys. }{\bf #1} (19#3) #2}
\def\cpc#1#2#3{
        {\it Comput. Phys. Commun. }{\bf #1} (19#3) #2}
\def\ijmp#1#2#3{
        {\it Int .J. Mod. Phys. }{\bf #1} (19#3) #2}
\def\ibid#1#2#3{
        {\it ibid }{\bf #1} (19#3) #2}
\def\jmp#1#2#3{
        {\it J. Math. Phys. }{\bf #1} (19#3) #2}
\def\jetp#1#2#3{
        {\it JETP Sov. Phys. }{\bf #1} (19#3) #2}
\def\ib#1#2#3{
        {\it ibid. }{\bf #1} (19#3) #2}
\def\mpl#1#2#3{
        {\it Mod. Phys. Lett. }{\bf #1} (19#3) #2}
\def\nat#1#2#3{
        {\it Nature (London) }{\bf #1} (19#3) #2}
\def\np#1#2#3{
        {\it Nucl. Phys. }{\bf #1} (19#3) #2}
\def\npsup#1#2#3{
        {\it Nucl. Phys. Proc. Sup. }{\bf #1} (19#3) #2}
\def\pl#1#2#3{
        {\it Phys. Lett. }{\bf #1} (19#3) #2}
\def\pr#1#2#3{
        {\it Phys. Rev. }{\bf #1} (19#3) #2}
\def\prep#1#2#3{
        {\it Phys. Rep. }{\bf #1} (19#3) #2}
\def\prl#1#2#3{
        {\it Phys. Rev. Lett. }{\bf #1} (19#3) #2}
\def\physica#1#2#3{
        { Physica }{\bf #1} (19#3) #2}
\def\rmp#1#2#3{
        {\it Rev. Mod. Phys. }{\bf #1} (19#3) #2}
\def\sj#1#2#3{
        {\it Sov. J. Nucl. Phys. }{\bf #1} (19#3) #2}
\def\zp#1#2#3{
        {\it Z. Phys. }{\bf #1} (19#3) #2}
\def\tmf#1#2#3{
        {\it Theor. Math. Phys. }{\bf #1} (19#3) #2}

\def\preprint{{\it preprint}}

\begin{flushright}
{\large ETH--TH/97--18}\\
{\rm August 1997\hspace*{.5 truecm}}\\
\end{flushright}

\vspace*{\fill}

\begin{center}
{\Large \bf 
Two-loop anomalous dimension in light-cone gauge with Mandelstam-Leibbrandt
prescription}
\\[0.5cm]
{\large 
Gudrun Heinrich and 
Zoltan Kunszt}\footnote{E-mails:
 gudrun@itp.phys.ethz.ch; kunszt@itp.phys.ethz.ch
 }\\[0.15 cm]
{\it  Institute of Theoretical Physics, ETH, Z\"urich, Switzerland}\\[0.15cm]
\end{center}
\vspace*{\fill}
\begin{abstract}
{\noindent\small 
All  the next-to-leading order results
 on  Altarelli-Parisi splitting functions
 have been  obtained in the literature either by using the  
operator product expansion method or by making use of the Curci Furmanski
Petronzio (CFP) formalism in conjunction with light-like axial gauge,
 principal value (PV) prescription and dimensional re\-gularization.
In this paper we present the calculation of
some non-singlet two-loop anomalous dimensions within the CFP formalism 
using light-cone axial gauge with Mandelstam-Leibbrandt (ML) prescription.
We make a detailed comparison between the intermediate results
given by the  (PV) versus the  (ML) method. 
We point out that the (ML) method is completely
consistent and avoids the ``phenomenological rules'' used in the case
of (PV) regularization.

}
\end{abstract}
\vspace*{\fill}

\newpage
\pagenumbering{arabic}

\section{Introduction}
At present and future high energy colliders a number of important
hard scattering cross sections can be measured with
high accuracy. The satisfactory description of these precision
data requires the evaluation of  next-to-leading order (NLO)
and in few cases even  next-to-next-to leading order (NNLO) corrections in
perturbative QCD. A number of NLO
corrections  are by now available in the literature 
(see e.g.~\cite{ESW} and references therein).  
NNLO corrections, however, could be evaluated only in few cases.  
The special technique
developed for the  calculation of the NNNLO corrections to 
the  total cross section of electron positron annihilation into
hadrons~\cite{gorkatlar}
has  recently been applied by Larin and Vermaseren~\cite{larverm} to get
the  NNLO and NNNLO anomalous dimensions for those operators
which contribute to the Bjorken sum rule and the Gross-Llewellyn Smith
sum rule.

In the case of  hard processes with two initial hadrons, however,
no complete NNLO result is available.
 Although the NNLO coefficient functions of the Drell-Yan process
have been obtained  in ref.~\cite{MvNeerven}, the 
phenomenological
application of this result requires the calculation  
 of the NNLO Altarelli-Parisi splitting functions (three-loop
anomalous dimensions) as well. It seems that  significant technical
 development has to be achieved before this  calculation can be
 carried out.

Two rather different methods~\cite{FRS},~\cite{cfp,cfpsinglet} have been used 
for the evaluation of the spin independent
two-loop anomalous dimensions. Both methods recently also have  been
 applied to the successful
calculation of  the NLO corrections of the spin dependent 
Altarelli-Parisi splitting functions ~\cite{vanne,
vogel1}.\footnote{A third method proposed 
in ref.~\cite{CSeik} makes use of the 
 the operator definition of the parton number densities and requires
the evaluation of  Feynam diagrams 
with attached eikonal factors. The method has not
yet been tested in higher order calculations.}
In the case  of the first   method~\cite{FRS} (OPE) 
 one has to evaluate the overall ultraviolet divergences
of twist-two local operator insertions.  The
results are obtained  in  moment space and the calculation can be carried
  out in Feynman gauge.
Unfortunately, the number of the operator insertions 
increases very rapidly in higher
orders and the treatment of operator mixing becomes increasingly 
cumbersome~\cite{collinsmix}.
 The second method~\cite{cfp,cfpsinglet} (CFP) 
 is based on the observation that
in axial gauge the two-particle irreducible kernel of the ladder
diagrams is finite. Therefore, using 
 renormalization group properties,
the anomalous dimensions are  given by some  projection of this
kernel in agreement with
 the factorization theorem of mass singularities~\cite{egmpr}.
The corrections to
the Altarelli-Parisi splitting functions are obtained directly in
configuration space by evaluating few Feynman diagrams. A  detailed
documentation of the NLO calculation using the CFP method
 appeared recently in ref.~\cite{ev}. 
A theoretically questionable
feature of the (CFP) method
is  the use of light-like axial gauge with principal value prescription.
It is not clear whether the PV prescription remains valid in 
higher orders since  standard quantization 
procedures in light-like axial gauge lead to the so-called
Mandelstam-Leibbrandt (ML) prescription~\cite{mand,leib}.
Another  technically unpleasant
feature of the method is that the 
 individual diagrams contributing to the finite
2PI  kernel  have  soft and collinear
singularities (which cancel in the sum).
 The actual algebraic complexity
of the CFP method  nevertheless is significantly 
smaller than the one of the OPE
method, motivating the further study of its technical aspects.
In particular,
it is of interest to investigate the CFP method with ML prescription.
A first attempt already appeared in the literature:  The one-loop
Altarelli-Parisi  splitting functions have been calculated  by Bassetto~\cite{Bas3}.
In the NLO calculation
with  ML prescription, the treatment of the ultraviolet part becomes
more consistent. This is a clear principal advantage. However,
the ML prescription is not without difficulties.
Firstly,  using  two light-like vectors the
integrals become more complicated. Secondly,  
 the singular $1/nq$ factor appearing
in the expression for the axial gauge propagator is regularized
with a $+i\ep$ prescription which results in   new so-called axial ghost 
contributions. 

\medskip

In this paper we  apply the CFP method
with ML prescription to the calculation of non-singlet
  two-loop anomalous dimensions. In particular, we calculate 
all terms which are proportional to $C_F^2$. An interesting
new feature of our calculation is  the evaluation of 
the axial ghost contributions.
In Section  2  we give a short review of the CFP formalism, in Section 3
 the PV and  ML prescriptions are discussed.     
Section 4 describes details of the calculation, treating separately the
virtual and the real contributions. In Section 5  we 
make a crtitical 
comparison between the results obtained by using the PV and ML prescription
and draw our conclusions from this discussion. 
 An Appendix contains
virtual integrals and phase space parametrizations.

\section{Framework of the calculation}\label{fn}

According to the factorization theorem,  simple
hard scattering cross sections can be written 
in perturbative QCD as convolutions
of finite hard scattering cross sections and singular parton number 
densities
\be
\sigma(\frac{Q^2}{\mu^2},\as(\mu^2),\ep)=
\hat{\sigma}(\frac{Q^2}{\mu^2},\as(\mu^2))\otimes
\Gamma(\as(\mu^2),\ep)\label{pQCDhard}
\ee
where the symbol $\otimes$ indicates the convolution over longitudinal 
momentum fractions which in moment space is reduced to a simple
product
\be 
\sigma_N(\frac{Q^2}{\mu^2},\as(\mu^2),\ep)=
\hat{\sigma}_N(\frac{Q^2}{\mu^2},\as(\mu^2))
\Gamma_N(\as(\mu^2),\ep)
\ee
where
\be
f_N=\int_0^1dx x^{N-1}f(x)\,.
\ee
Our discussion will be  valid for the non-singlet parton densities,
but it can be  easily   generalized to the singlet case (matrix problem).
The physical cross section $\sigma_H $ must be independent from $\mu^2$,
 therefore the deviation of the physical dimension 
of $\hat{\sigma}$   from its canonical value (anomalous dimension)
\be\label{anomsig}
\gamma_N(\as(\mu^2))=-
\frac{d}{d\ln \mu^2}\ln\hat\sigma_N(\frac{Q^2}{\mu^2},\as(\mu^2))
\ee
is related to the moments of parton number densities 
\be
\gamma_N(\as(\mu^2)=\frac{d}{d\ln \mu^2}\ln
\Gamma_N(\as(\mu^2),\ep)
=\beta(\as,\ep)\frac{d}{d\as}\ln \Gamma_N(\as(\mu^2),\ep)
\label{gamep}
\ee
where in the $\overline{\rm MS}$ scheme 
\be
\beta(\as,\ep)=\frac{d\as}{d\ln \mu^2}=-\ep\as+\beta(\as)\,.
\ee 
Expanding Eq.~(\ref{gamep}) in $\ep$ and
keeping only the leading order term one gets
\be 
\gamma_N(\as(\mu^2) = -\frac{d}{d\ln\as}\Gamma^{(1)}_N(\as)
\label{gamGamN}
\ee
where $\Gamma_N^{(1)}$ is the coefficient of $1/\ep$  in the expansion
of $\Gamma_N(\as(\mu^2),\ep)$
\be
\Gamma_N(\as(\mu^2),\ep) = 1 + \sum_i
\frac{\Gamma^{(i)}_N(\as)}{\ep^i}\,.
\label{singlepoleG}
\ee 
In the parton picture, 
the  physical hard scattering 
cross section $\sigma_H$ with incoming hadron $H$
 is obtained by convoluting the
partonic cross section $\hat{\sigma}\otimes\Gamma$ 
with "bare" parton densities $q^{B,H}$. The finite
quantity $(\Gamma\otimes q^{B,H})(x)$ $=$ $q_{f/H}(x,\mu^2)$ 
is then interpreted as the physical (renormalized)
number density in the longitudinal momentum fraction $x$ of parton  type $f$
in hadron 
$H$.   This factorization scheme is shown diagrammatically in Figure 1.

\begin{center}
\begin{picture}(100,130)(40,-20)
\Text(0,50)[]{$\sigma_H$}
\CArc(0,50)(15,0,360)
\Line(-6,20)(-6,37)
\Line(-4,20)(-4,35)
\Line(4,20)(4,35)
\Line(6,20)(6,37)
\DashLine(-15,80)(-10,60){3}
\DashLine(15,80)(10,60){3}
\Line(40,50)(45,50)
\Line(40,54)(45,54)
\Text(90,80)[]{$\hat{\sigma}$}
\Text(90,45)[]{$\Gamma$}
\Text(90,10)[]{$q_{B,H}$}
\Boxc(90,80)(25,15)
\Boxc(90,45)(25,15)
\Boxc(90,10)(25,15)
\Line(84,2)(84,-10)
\Line(86,2)(86,-10)
\Line(94,2)(94,-10)
\Line(96,2)(96,-10)
\DashLine(75,102)(80,90){3}
\DashLine(105,102)(100,90){3}
\CArc(90,77)(10,210,330)
\CArc(90,43)(10,210,330)
\Vertex(90,62){2}
\Vertex(90,25){2}
\Line(90,62)(79,52)
\Line(90,62)(101,52)
\Line(90,25)(79,17)
\Line(90,25)(101,17)
\Line(120,50)(125,50)
\Line(120,54)(125,54)
\DashLine(145,95)(150,80){3}
\DashLine(175,95)(170,80){3}
\Text(160,70)[]{$\hat{\sigma}$}
\Boxc(160,70)(25,15)
\CArc(160,67)(10,210,330)
\Text(160,30)[]{$q_{f/H}$}
\Boxc(160,30)(25,15)
\Line(154,22)(154,10)
\Line(156,22)(156,10)
\Line(164,22)(164,10)
\Line(166,22)(166,10)
\Vertex(160,50){2}
\Line(160,50)(149,37)
\Line(160,50)(171,37)
\Text(60,-25)[]{Figure 1: Factorization scheme}
\end{picture}
\end{center}

\noindent
The parton evolution equation in moment space becomes simply 
\be
\frac{dq_N(\mu^2)}{d\ln \mu^2}=\gamma_N(\as) q_N(\mu^2)\,,
\ee
and therefore  the anomalous dimensions $\gamma_N(\as)$ 
defined through Eq.~(\ref{anomsig}) can be
interpreted as  the moments
of the Altarelli-Parisi splitting function
\be
\gamma_{N,q/q}(\as)=\int_0^1dx x^{N-1} P_{q/q}(x)
 \ee
where we consider the density of a quark within a  quark.
These relations allow to calculate $P_{q/q}(x,\as) $ in a power series
in $\as$
\be
P_{q/q}(x,\as)=\left(\frac{\as}{2\pi}\right) P^{(0)}_{q/q}(x)+
\left(\frac{\as}{2\pi}\right)^2P^{(1)}_{q/q}(x) + {\cal O}(\as^3)
\ee
which results in the following expansion of  $\Gamma_{q/q}(x,\as,\ep)$ 
\be
\Gamma_{q/q}(x,\as,\ep)=\delta(1-x)-\frac{1}{\ep}
\left\{\left(\frac{\as}{2\pi}\right) P^{(0)}_{q/q}(x)
+\frac{1}{2}\left(\frac{\as}{2\pi}\right)^2P^{(1)}_{q/q}(x)
+\ldots\right\}
 + {\cal O}\left(\frac{1}{\ep^2}\right).
\ee
Here 
$P_{q/q}^{(0)}(x)$ is the first order
Altarelli-Parisi splitting function:
$$P_{q/q}^{(0)}(x)=C_F\,\left(\frac{1+x^2}{1-x}\right)_+$$

\noindent
The CFP method for calculating $\Gamma_{q/q}(x,\alpha_S,\epsilon)$
in terms of a well-defined projection
of   the 2PI kernel $K_{0,aa^{\prime}}^{bb^{\prime}}(k,k')$ (upper lines are
non-amputed, lower lines are amputed)  of the  ladder diagram
\be 
\Gamma=\frac{1}{1-{\cal P}K}, \quad K=K_0 \left(1-(1-{\cal P})K_0\right)^{-1}
\ee
consists in using a projector ${\cal P} =
{\cal P}_{\epsilon}\otimes {\cal P}_n$ which can be considered
as a pinching of  the rungs of the ladder.
This symbolic representation of the projector  ${\cal P}$
indicates a projection
  acting on the $\ep-$dependence (${\cal P}_{\epsilon}$)
 and a projection acting
on the spin, colour and momentum dependence ${\cal P}_n$ of the kernel.
The rungs   above the pinch are put on-shell and considered as external
on-shell legs for the upper part of the diagram with
 averaging over spin and   colour labels.
The  rungs below the pinch remain  non-amputed. They have to be summed
over the spin and colour indices
in conjunction with the  insertion of  the operator
  $\frac{\not n}{4kn}$ for quarks and  $-g_{\mu\nu}$ for gluons
and have to be integrated over the off-shell loop momentum
$k$. Finally, the projector extracts  the single pole term of the
whole expression, setting
$\ep$ to zero in its residuum according to Eq.~(\ref{singlepoleG}).
The singular non-singlet quark  density 
is therefore given  by the expression
\begin{eqnarray}
\Gamma_{q/q}(x,\alpha_S,\frac{1}{\epsilon})&=&Z_{F}
\Biggl\{\delta(1-x) \nonumber \\ 
&& + PP\big\{x\int\frac{d^mk}{(2\pi)^m}\delta(x-\frac{kn}{pn}){\rm
    Trace}\left[\frac{\not n}{4kn}
\frac{K}{1-{\cal P}K}\not p\right]\big\}\Biggr\}\label{9a}
\end{eqnarray}
where $PP$ means to take the pole part. In fixed order
perturbation theory we can expand $\Gamma-1= K/(1-{\cal P}K)$
in terms of  the  2PI kernel $K_0$  as 
\begin{equation}
\Gamma-1={\cal P}K_0+{\cal P}(K_0)^2-{\cal P}(K_0{\cal
P}K_0)+\ldots \label{K0exp}
\end{equation}
These equations are the basic ones to calculate the anomalous
dimensions in higher orders. In two-loop order
the contributions  can be classified into virtual
and real contributions. The virtual contributions are
diagrams where only one internal parton line is cut and
the diagrams have virtual subdiagrams. The real contributions
are defined in terms of diagrams where two parton lines are cut and
without any virtual subdiagram. In the CFP method
it is convenient to treat separately the contribution from $Z_F$, the
renormalization
factor of the external leg, and hence to introduce the notation
\begin{eqnarray}
\Gamma_{q/q}(x,\alpha_S,\frac{1}{\epsilon})&=&Z_F\,\hat{\Gamma}_{q/q}(x,\alpha_S,\frac{1}{\epsilon})\nonumber\\
\hat{\Gamma}_{q/q}(x,\alpha_S,\frac{1}{\epsilon})&=&\delta(1-x)\nonumber\\
&&-\frac{1}{\epsilon}\,\left\{\left(\frac{\alpha_S}{2\pi}\right)\,
\hat{P}^{(0)}_{q/q}(x)+\frac{1}{2}\left(\frac{\alpha_S}{2\pi}\right)^2\,
\hat{P}^{(1)}_{q/q}(x)\,+\,{\cal O}(\alpha_S^3)\right\}\,+\,{\cal
O}(\frac{1}{\epsilon^2})\label{gamahat}\\
Z_F&=&1-\frac{1}{\epsilon}\,\xi(\alpha_S)\,+\,{\cal
  O}(\frac{1}{\epsilon^2})
\nonumber\\
P_{q/q}(x,\alpha_S)&=&\hat{P}_{q/q}(x,\alpha_S)+\xi(\alpha_S)\,\delta(1-x).\label{hatp}
\end{eqnarray}
An important simplification in the 
the CFP method is that  one has to
calculate  only   $\hat{P}^{(1)}_{a/b}(x)$
 because the contribution of the wave function renormalization factor   
 $Z_F$   at $x=1$  can easily be obtained
 from fermion number conservation and momentum conservation sum rules.

\section{Principal value versus Mandelstam-Leibbrandt prescription}
\medskip\noindent
As we already noted,
a crucial ingredient of the CFP method  is the use of the light-cone
gauge (or light-like axial gauge), defined by 
\begin{equation}
n^{\mu}A_{\mu}^a(x)=0\quad;\quad n^2 = 0 \label{gauge}
\end{equation}
which formally leads to a gluon propagator of the form 
\begin{equation}
D_{\mu\nu}^{ab}(q) = \frac{-i\delta^{ab}}{q^2+i\ep}\{g_{\mu\nu}-\frac{n_{\mu}q_{\nu}+n_{\nu}q_{\mu}}{qn}\}\label{axprop}\,.
\end{equation}
The $1/qn-$factor in the gluon propagator gives rise to the so-called
``spurious poles'', singular terms
in both the real and virtual contributions. Although these singular
contributions must cancel in  gauge invariant quantities, one has to use 
some regularization method
 for the evaluation of the individual contributions.
As yet, in all calculations based on the CFP method, the 
principal value (PV) prescription has been  used which is defined as
\begin{eqnarray}
\frac{1}{qn}&\to& \lim_{\delta\to 0}\frac{1}{2}\left(\frac{1}{qn+i\delta (pn)}+\frac{1}{qn-i\delta
(pn)}\right)=\frac{qn}{(qn)^2+\delta^2(pn)^2}\label{pv}\\
&&\nonumber\\
\mbox{where}&&p^2=0\,;\,pn\not =0\,;\,\vec{p}_T=\vec 0\,.\nonumber
\end{eqnarray}
It is well-known\footnote{The subject of
quantization and renormalization  in noncovariant gauges and their use in
perturbation theory is described in refs.~\cite{leibbrandt,Basbook,Leibbook}.}
however that the PV prescription is incompatible with   
Wick rotation and hence power counting breaks down. Nevertheless, 
the spurious poles appearing as $\ln\delta,\,\ln^2\delta-$terms 
in individual contributions cancel
 in the sum of all virtual and all real contributions.
Another difficulty of the PV method is related to the ultraviolet
renormalization constants which become  dependent on
longitudinal momentum fractions and on $\ln\delta$.
CFP performed
the ultraviolet renormalization by 
subtracting  all ultraviolet poles,  normal and spurious ones, calling this
procedure ``a phenomenological rule''.

After the work of CFP it was pointed out that 
 the principal value prescription is not consistent
with  canonical quantization  in light-like axial
gauge~\cite{Bas2,cartor}. 
Correctly performed canonical quantization  leads to
the ML  prescription defined as 
\begin{equation}
\frac{1}{qn}\to \lim_{\eta\to 0^+}\frac{qn^*}{qn\,qn^*+i\eta}\label{ml1}
\end{equation}
where the vector $n_{\mu}^*$ is ``dual'' to $n_{\mu}$:
\be
n_{\mu}=(n_0,\vec n)\,;\,n_{\mu}^*=(n_0,-\vec n)\,;\,(n^*)^2 = 0.
\ee
The ML prescription also appears in the literature as 
\be
\frac{1}{qn}\to \lim_{\eta\to
0^+}\frac{1}{qn+i\eta\,{\rm sign}(qn^*)}\label{ml3}.
\ee
In the present calculation, these alternative definitions
 lead to the same
result. 
A crucial property of the ML prescription is that the spurious poles are
placed in the complex $q_0-$plane in the same way as the ``usual'' covariant
poles. Therefore, Wick rotation and  power counting theorems for UV
divergences remain valid. The propagator in the ML prescription
can be decomposed into a term corresponding  to the propagation of the
physical polarizations and into a term which describes the
propagation of scalar and longitudinal gluons in the $qn=0$ plane
\be
<0|T\{A^a_{\mu}(x)A^b_{\nu}(0)\}|0>=
<0|T\{T^a_{\mu}(x)T^b_{\nu}(0)\}|0> + <0|T\{L^a_{\mu}(x)L^b_{\nu}(0)\}|0>
\ee
where
\begin{eqnarray}
<0|T\{T^a_{\mu}(x)T^b_{\nu}(0)\}|0>&=&
\frac{i\delta^{ab}}{(2\pi)^4}\int\frac{d^4q\,e^{iqx}}{q^2+i\ep}
\nonumber\\  &&
\left(-g_{\mu\nu}
+\frac{(n_{\mu}q_{\nu}+q_{\mu}n_{\nu})}{q^2_{\perp}}\,\frac{2qn^*}{nn^*}
-\frac{n_{\mu}g_{0\nu}+n_{\nu}g_{0\mu}}{q^2_{\perp}}\,q^2\right)
\end{eqnarray}
and
\begin{eqnarray}
<0|T\{L^a_{\mu}(x)L^b_{\nu}(0)\}|0>&=&
-\frac{i\delta^{ab}}{(2\pi)^4}\int\frac{d^4q\,e^{iqx}}{q^2+q^2_{\perp}+i\ep}\nonumber\\
&&\left(\frac{(n_{\mu}q_{\nu}+q_{\mu}n_{\nu})}{q^2_{\perp}}\,\frac{2qn^*}{nn^*}
-\frac{n_{\mu}g_{0\nu}+n_{\nu}g_{0\mu}}{q^2_{\perp}}\,(q^2+q^2_{\perp})\right)
\,.
\end{eqnarray}
Adding up these contributions and using $q^2+q^2_{\perp}=2q^+q^-=2(qn^*)(qn)/nn^*$
we get the  axial-gauge propagator
with  ML regularization
\be\label{MLprop}
D^{ab}_{\mu\nu}(x)=\frac{i\delta^{ab}}{(2\pi)^4}\int
\frac{d^4q\,e^{iqx}}{q^2+i\ep}
\left(-g_{\mu\nu}+\frac{(n_{\mu}q_{\nu}+q_{\mu}n_{\nu})\,qn^*}
{qn\,qn^*+i\ep}\right).
\ee
The discontinuity of this propagator can be decomposed into the
physical axial-gauge contribution and  an unphysical
contribution 
\bea
disc\Big[\frac{id_{\mu\nu}(q)}{q^2}\Big]&=&
2\pi\theta(q_0)
\left\{-g_{\mu\nu}
+\frac{2qn^*}{nn^*}\,
\frac{(n_{\mu}q_{\nu}+n_{\nu}q_{\mu})}
{q_{\bot}^2}\right\}\delta(q^2)\nonumber\\ &&
-2\pi\theta(q_0)\left\{\frac{2qn^*}{nn^*}\,
\frac{(n_{\mu}q_{\nu}+n_{\nu}q_{\mu})}
{q_{\bot}^2}\right\}\delta(q^2+q_{\bot}^2)\,.
\label{disprop}
\eea
The term proportional
to the delta function $\delta(q^2+q^2_{\perp})$ is the 
so-called axial ghost contribution. It has been shown by Bassetto
\etal~\cite{Bas2}
that all vectors of the physical Hilbert-space are 
annihilated by the creation operator
of these degrees of freedom. Therefore, similarly to the
Gupta-Bleuler ghosts of QED, they decouple from  
the S-matrix. The ghosts have negative  mass squared, indefinite
metric, they live in the $qn=0$ plane
and their polarization sum is
\be 
\sum_{\lambda=1,2}e^{\lambda}_{\mu}(q)e^{\lambda}_{\nu}(q)=-\frac{2qn^*}{nn^*}\,\frac{(n_{\mu}q_{\nu}+q_{\mu}n_{\nu})}{q^2_{\perp}}
\,.
\ee

In the ML prescription 
there is a cancellation between the
standard axial gauge contribution  and the axial ghost
contribution: In the  limit $q^2_{\perp}\to 0$
the spurious poles $1/q^2_{\perp}$
appearing in the denominator of the
polarization sums cancel each other as can be seen from Eq.~(\ref{disprop}).
As a consequence, when  contributions of diagrams 
with two cut gluon lines are calculated, the result must be independent from
the regularization scheme of these spurious
poles.  One may  use for example dimensional regularization
or a PV regularization. The PV regularization can be chosen such that the
standard axial gauge contribution from the first term
 in Eq.~(\ref{disprop}) becomes identical to the one of the PV
calculation of CFP where no ghosts are included at all. 
Choosing the same  PV
regularization for the ghost part, one finds that the spurious poles cancel
within the sum of all ghost diagrams, but a finite contribution is remaining
from the ghost diagrams. 

Using only dimensional regularization, mixed products of spurious, soft
and collinear poles are obtained  and one cannot see the details of the
cancellation mechanism. In addition, one deals with poles
 of higher powers such that the integrals
have to be calculated in higher order in the regularization 
parameter $\ep$. If we choose
 PV regularization, however, the
various (spurious, soft, collinear) 
singular contributions can be clearly separated 
from each other. In particular, the result we get with ML prescription for the
real contributions differs from the CFP result only because of the axial ghost
contributions.

One can find a similar decomposition also for the  virtual contributions
 with the help of 
a formula   found  in refs.~\cite{oegren,capper}.
Let  us consider the integral
\be
J^{A}_n(k_1\ldots k_{n-1})=\int d^m q \frac{1}{q^2\,(q+k_1)^2\ldots (q+k_{n-1})^2\,qn}
\ee
defined in Appendix A. Using exponential parametrization for the propagator
denominator factors and the ML prescription to regulate the $1/qn-$denominator
leads to 
\bea
J^{A,{\rm ML}}_n(k_1\ldots
k_{n-1})&=&\frac{1}{i^n}\int_0^{\infty}da_0\ldots da_{n-1}\,\exp\{i\sum
a_jk_j^2-iR^2/z-z\ep\}\nonumber\\
&&\cdot\int d^m q \frac{\exp\{iz\,q^2\}}{q^+-R^+/z + i\eta\,{\rm
sign}\,(q^--R^-/z)}\\
&&\nonumber\\
R^{\mu}&=&a_1k_1^{\mu}+\ldots+a_{n-1}k_{n-1}^{\mu}\nonumber\\
z&=&a_0+\ldots +a_{n-1}\,.\nonumber
\eea
The $q-$integral $$J^{ML}_q=\int dq^+dq^-d^{(m-2)}\vec
q_{\perp}\,\frac{\exp\{iz\,q^2\}}{q^+-R^+/z + i\eta\,{\rm sign}(q^--R^-/z)}$$
has been evaluated using the Cauchy theorem and integrating first over $dq^+$,
then over $dq^-$. The integral over the transverse momenta  is a  standard
Gaussian integral, and one obtains finally
\be
J^{ML}_q=\frac{-\pi^{\frac{m}{2}}(iz)^{1-\frac{m}{2}}}{R^+ + i\eta\, z\,{\rm sign}\,R^-}\Big\{1-\exp\{2iR^+R^-/z-2\eta|R^-|\}\Big\}
\ee
leading to 
\bea
J^{A,{\rm ML}}_n(k_1\ldots
k_{n-1})&=&\frac{-\pi^{\frac{m}{2}}}{i^{n+\frac{m}{2}-1}}\int_0^{\infty}da_0\ldots
da_{n-1}\,\exp\{i\sum
a_jk_j^2-z\ep\}\cdot z^{1-\frac{m}{2}}\nonumber\\
&&\cdot \frac{\exp\{-i\,R^2/z\}}{R^+ + i\eta\,z\,{\rm
sign}\,R^-}\Big\{1-\exp\{2iR^+R^-/z-2\eta|R^-|\}\Big\}\label{virtML}
\eea
whereas PV regularization leads to 
\bea
J^{A,{\rm PV}}_n(k_1\ldots
k_{n-1})&=&\frac{-\pi^{\frac{m}{2}}}{i^{n+\frac{m}{2}-1}}\int_0^{\infty}da_0\ldots
da_{n-1}\,\exp\{i\sum
a_jk_j^2-z\ep\}\cdot z^{1-\frac{m}{2}}\nonumber\\
&&\cdot \exp\{-i\,R^2/z\}\, \frac{R^+}{(R^+)^2 + z^2\,\delta^2(p^+)^2}\label{virtPV}
\eea
We can see from Eq.~(\ref{virtML}) that  in the  ML scheme
 there is a cancellation between 
 the first and the second term as  $R_+\to 0$.
Therefore,  the result will have no spurious poles and will be independent
from the regularization method of the spurious singularity.
Similar to  the case of the real contributions,
it is useful  
to  use   PV regularization for the individual spurious singularities even
{\it within} the ML scheme.
We emphasize that the application of a PV regularization here
has nothing to do with the prescription required {\it before} carrying out
the loop integral. Using PV regularization before doing the $d^m q-$integral 
leads to the result (\ref{virtPV}). \\
Comparing Eqs.~(\ref{virtML}) and (\ref{virtPV}) shows that the PV result can
be identified as a subpart of the ML result by dropping the second term in
Eq.~(\ref{virtML}) and taking the real part of the $1/(R^+ + i\eta\,z\,{\rm
sign}R^-)$ term. Hence the difference between
the results for the virtual parts in the PV and ML prescription  is given
completely by the contributions from the second term in Eq.~(\ref{virtML}).
It follows that
the difference between the sum  of all PV- and  all ML-regulated virtual
contributions must be equal to the sum of the real 
axial ghost contributions. That is indeed  what we will find
as a result of  explicit calculation.

\section{Details of the calculation}
\subsection{Diagrams contributing to terms with colour factor $C_F^2$}
 Since every colour structure defines a gauge invariant contribution
we can simplify our exploratory study of the application of the
ML prescription by  considering only the terms that are
 proportional to $C_F^2$.
Besides the obvious advantage of reducing  the number of 
contributing Feynman diagrams,  
this colour structure has two additional simplifying features:
i) the ultraviolet counter term is vanishing, ii) the sums of the
real and virtual contributions are separately free from soft
and collinear singularities. The axial-ghost contributions, however,
remain important and therefore
 the $C_F^2$ structure gives a good opportunity to study
these contributions in isolation from other complications.
 The contributing Feynman diagrams are listed
in Fig.~2. The axial ghosts are denoted by dotted lines replacing cut gluon
lines. 

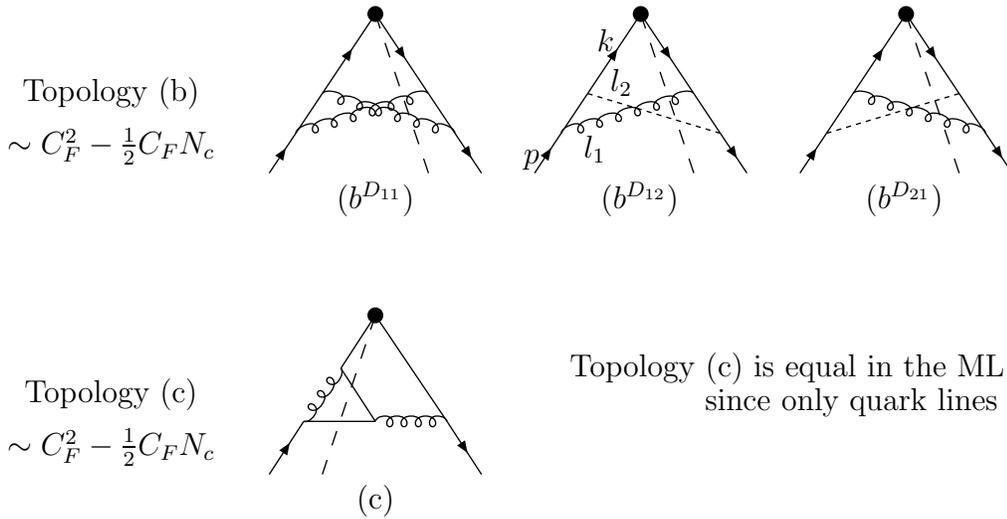
\begin{center}
\begin{picture}(200,50)(0,0)
\Text(-100,-30)[]{Topology (b)}
\Text(-100,-50)[]{$\sim C_F^2-\frac{1}{2}C_FN_c$}
\Text(0,-70)[]{$(b^{D_{11}})$}
\DashLine(0,0)(20,-60){6}
\ArrowLine(-20,-30)(0,0)
\ArrowLine(0,0)(20,-30)
\Vertex(0,0){3}
\Gluon(-20,-30)(30,-45){2}{6}
\Line(-20,-30)(-30,-45)
\Line(20,-30)(30,-45)
\Gluon(-30,-45)(20,-30){2}{6}
\ArrowLine(-40,-60)(-30,-45)
\ArrowLine(30,-45)(40,-60)
\Text(100,-70)[]{$(b^{D_{12}})$}
\Text(83,-51)[]{$l_1$}
\Text(93,-26)[]{$l_2$}
\Text(59,-55)[]{$p$}
\Text(87,-10)[]{$k$}
\DashLine(100,0)(120,-60){6}
\ArrowLine(80,-30)(100,0)
\ArrowLine(100,0)(120,-30)
\Vertex(100,0){3}
\DashLine(80,-30)(130,-45){2}
\Line(80,-30)(70,-45)
\Line(120,-30)(130,-45)
\Gluon(70,-45)(120,-30){2}{6}
\ArrowLine(60,-60)(70,-45)
\ArrowLine(130,-45)(140,-60)
\Text(200,-70)[]{$(b^{D_{21}})$}
\DashLine(200,0)(220,-60){6}
\ArrowLine(180,-30)(200,0)
\ArrowLine(200,0)(220,-30)
\Vertex(200,0){3}
\Gluon(180,-30)(230,-45){2}{6}
\Line(180,-30)(170,-45)
\Line(220,-30)(230,-45)
\DashLine(170,-45)(220,-30){2}
\ArrowLine(160,-60)(170,-45)
\ArrowLine(230,-45)(240,-60)
\end{picture}
\end{center}

\begin{center}
\begin{picture}(200,100)(0,0)
\Text(-100,-30)[]{Topology (c)}
\Text(-100,-50)[]{$\sim C_F^2-\frac{1}{2}C_FN_c$}
\Text(0,-70)[]{(c)}
\DashLine(0,0)(-20,-60){6}
\Line(-13,-20)(0,0)
\Line(0,0)(27,-40)
\ArrowLine(27,-40)(40,-60)
\Vertex(0,0){3}
\Line(-13,-20)(0,-40)
\Gluon(-13,-20)(-27,-40){2}{3}
\Line(-27,-40)(0,-40)
\Gluon(0,-40)(27,-40){2}{4}
\ArrowLine(-40,-60)(-27,-40)
\Text(200,-20)[]{Topology (c) is equal in the ML and PV schemes}
\Text(200,-33)[]{since only quark lines are cut}
\end{picture}
\end{center}

\newpage

\begin{center}
\begin{picture}(200,100)(0,0)
\Text(-100,-30)[]{Topology (d)}
\Text(-100,-50)[]{$\sim C_F^2-\frac{1}{2}C_FN_c$}
\Text(0,-70)[]{(d)}
\Text(-41,-55)[]{$p$}
\Text(-13,-10)[]{$k$}
\DashLine(0,0)(20,-60){6}
\Line(-13,-20)(0,0)
\Line(0,0)(27,-40)
\ArrowLine(27,-40)(40,-60)
\Vertex(0,0){3}
\Line(-13,-20)(0,-40)
\Gluon(-13,-20)(-27,-40){2}{3}
\Line(-27,-40)(0,-40)
\Gluon(0,-40)(27,-40){2}{4}
\ArrowLine(-40,-60)(-27,-40)
\Text(100,-30)[]{Topology (e)}
\Text(100,-50)[]{$\sim C_F^2$}
\Text(200,-70)[]{(e)}
\Text(159,-55)[]{$p$}
\Text(190,-26)[]{$k$}
\DashLine(200,0)(200,-60){6}
\Line(180,-30)(200,0)
\Line(200,0)(220,-30)
\Vertex(200,0){3}
\GlueArc(195,-27)(17,95,200){2}{4}
\Line(180,-30)(170,-45)
\Line(220,-30)(230,-45)
\Gluon(170,-45)(230,-45){2}{7}
\ArrowLine(160,-60)(170,-45)
\ArrowLine(230,-45)(240,-60)
\end{picture}
\end{center}

\begin{center}
\begin{picture}(200,100)(0,0)
\Text(-100,-30)[]{Topology (h)} 
\Text(-100,-50)[]{$\sim C_F^2$}
\Text(0,-70)[]{$(h^{D_{11}})$}
\DashLine(0,0)(0,-60){6}
\ArrowLine(-20,-30)(0,0)
\ArrowLine(0,0)(20,-30)
\Vertex(0,0){3}
\Gluon(-20,-30)(20,-30){2}{6}
\Line(-20,-30)(-30,-45)
\Line(20,-30)(30,-45)
\Gluon(-30,-45)(30,-45){2}{7}
\ArrowLine(-40,-60)(-30,-45)
\ArrowLine(30,-45)(40,-60)
\Text(100,-70)[]{$(h^{D_{12}})$}
\DashLine(100,0)(100,-60){6}
\ArrowLine(80,-30)(100,0)
\ArrowLine(100,0)(120,-30)
\Vertex(100,0){3}
\DashLine(80,-30)(120,-30){2}
\Line(80,-30)(70,-45)
\Line(120,-30)(130,-45)
\Gluon(70,-45)(130,-45){2}{7}
\ArrowLine(60,-60)(70,-45)
\ArrowLine(130,-45)(140,-60)
\Text(200,-70)[]{$(h^{D_{21}})$}
\DashLine(200,0)(200,-60){6}
\ArrowLine(180,-30)(200,0)
\ArrowLine(200,0)(220,-30)
\Vertex(200,0){3}
\Gluon(180,-30)(220,-30){2}{6}
\Line(180,-30)(170,-45)
\Line(220,-30)(230,-45)
\DashLine(170,-45)(230,-45){2}
\ArrowLine(160,-60)(170,-45)
\ArrowLine(230,-45)(240,-60)
\end{picture}
\end{center}
\begin{center}
\begin{picture}(200,100)(0,0)
\Text(-100,-30)[]{Topology (i)}
\Text(-100,-50)[]{$\sim C_F^2$}
\Text(0,-75)[]{$(i^{D_{11}})$}
\Text(100,-75)[]{$(i^{D_{21}})$}
\DashLine(0,0)(0,-60){6} 
\Vertex(0,0){3}
\CArc(0,-17)(17,0,180)
\CArc(0,-17)(17,180,0)
\Gluon(-17,-17)(17,-17){2}{4}
\Vertex(0,-34){3}
\Line(0,-34)(-25,-60)
\Line(25,-60)(0,-34)
\Gluon(-15,-50)(15,-50){2}{4}
\DashLine(100,0)(100,-60){6} 
\Vertex(100,0){3}
\CArc(100,-17)(17,0,180)
\CArc(100,-17)(17,180,0)
\Gluon(83,-17)(117,-17){2}{4}
\Vertex(100,-34){3}
\Line(100,-34)(75,-60)
\Line(125,-60)(100,-34)
\DashLine(85,-50)(115,-50){2}
\Text(200,-75)[]{$(i^{D_{12}})$}
\DashLine(200,0)(200,-60){6} 
\Vertex(200,0){3}
\CArc(200,-17)(17,0,180)
\CArc(200,-17)(17,180,0)
\DashLine(183,-17)(217,-17){2}
\Vertex(200,-34){3}
\Line(200,-34)(175,-60)
\Line(225,-60)(200,-34)
\Gluon(185,-50)(215,-50){2}{4}
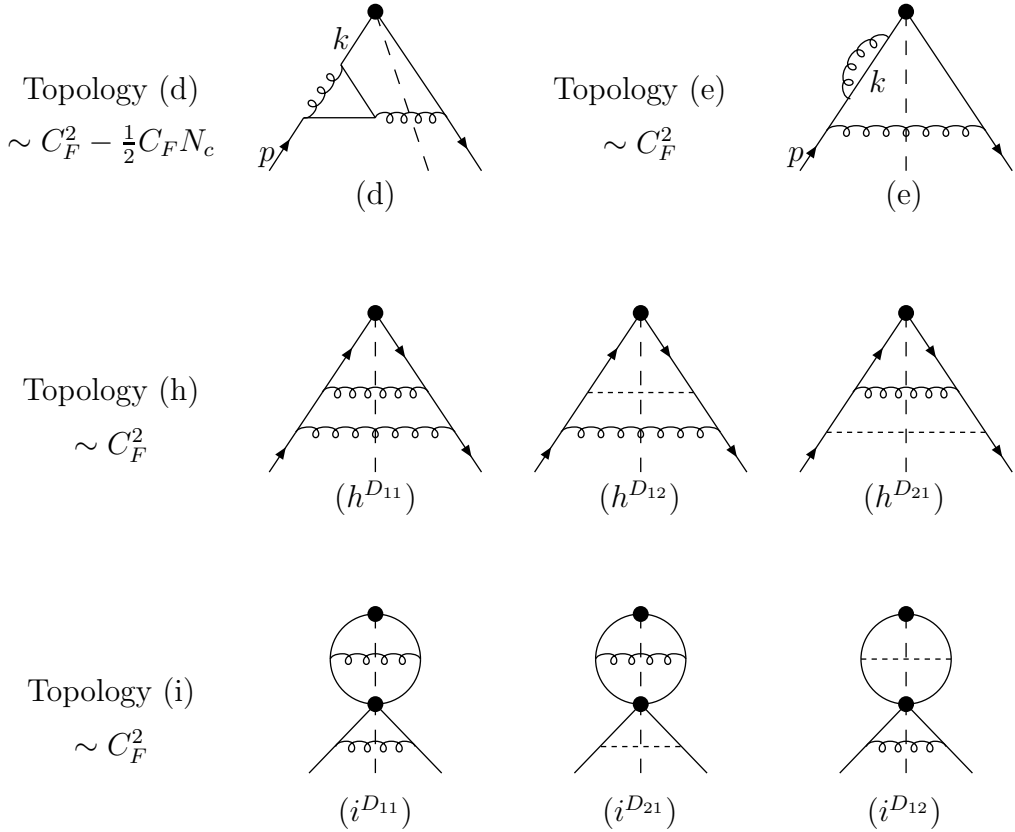
\Text(60,-100)[]{{Figure 2:  Diagrams contributing 
to $\hat \Gamma^{(1)}_{q/q}(x)$\, 
with colour factor $C_F^2$}}
\end{picture}
\end{center}

\vspace{3.5cm}

\subsection{Virtual corrections}
\subsubsection{Quark selfenergy}
The quark selfenergy (see topology (e) in Fig.\,2) 
can be split  into two terms
\bea
\Sigma &=& \Sigma^F +\Sigma^A \nonumber\\
&=&C_F\, g^2\int\frac{d^m q}{(2\pi)^m}\,\frac{\gamma_{\mu}\,(\not k + \not
q)\,\gamma_{\nu}}{(k+q)^2\,q^2}\left\{-g_{\mu\nu}+\frac{q_{\mu}n_{\nu}+q_{\nu}n_{\mu}}{qn}\right\}\label{si}
\eea
where 
$\Sigma^F$ denotes that part of the expression which stems from the Feynman
part (the part proportional to $-g_{\mu\nu}$) of the gluon propagator, 
$\Sigma^A$ denotes the remaining axial  part.
$\Sigma^F$ can be written in terms of simple loop integrals as 
\be
\Sigma^F =g^2 C_F \frac{(m-2)}{(2\pi)^m}\,\left(\not k \,J^F_2(k) +
\gamma^{\nu}J^F_{2\nu}(k)\right)\label{a2}
\ee
where 
\begin{eqnarray*}
J^F_2(k) &=& \int d^m q \frac{1}{q^2\,(q+k)^2}\quad=\quad Q^k_{\epsilon}\,T_0\quad;\quad T_0=\frac{1}{\epsilon_{UV}}+2\\
J^F_{2\mu}(k) &=&  \int d^m q \frac{q_{\mu}}{q^2\,(q+k)^2}\quad=\quad -Q^k_{\epsilon}\,T_1
k^{\mu}\quad;\quad T_1=\frac{1}{2}\,T_0\\
&&\nonumber\\
Q^k_{\epsilon} &=& i\pi^{\frac{m}{2}}\Gamma (1+\epsilon)(-k^2)^{-\epsilon}.
\end{eqnarray*}
We use the notation $\ep_{UV}$ for $\ep$ when the singular contribution comes
from the ultraviolet region. 
The axial part $\Sigma^A$ is given by 
\be
\Sigma^A 
=- g^2 C_F  \frac{1}{(2\pi)^m}\left(2 k^2\not n\,J_2^A(k) 
+ \not n\,J^A_{2\mu}(k)\gamma^{\mu}\not k + \not
k\,J^A_{2\mu}(k)\gamma^{\mu}\not n\right)\label{A6}
\ee
where
\bea
J_2^A(k)&=&
 \int d^m q \frac{1}{q^2\,(q+k)^2\,qn}\quad =\quad -\frac{Q^k_{\epsilon}}{kn}\,P_0(k)\nonumber\\
J^A_{2\mu}(k) &=&  \int d^m q \frac{q_{\mu}}{q^2\,(q+k)^2\,qn}\quad = \quad \frac{Q^k_{\epsilon}}{kn}\,(P_1(k) k^{\mu} + P_2(k) p^{\mu} +P_3(k)
n^{\mu})
\,.
\eea
The actual values of the form factors $P_0,P_1,P_2,P_3$
are different for the PV and  the ML prescription and can
be found in Appendix A.
In terms of form factors we obtain
\begin{eqnarray}
\Sigma^F &=&  C_F\,\frac{g^2}{(2\pi)^m}Q^k_{\epsilon}\frac{(m-2)}{2}\not k \,T_0 \\
\Sigma^A &=&-C_F \,\frac{g^2}{(2\pi)^m}\,\frac{Q^k_{\epsilon}}{kn}\left(2\not n
k^2\,(P_1(k)-P_0(k))+P_2(k)(\not n\not p\not k +\not k\not p\not n)\right)\nonumber\\
&=&-C_F \,\frac{g^2}{(2\pi)^m}\,\frac{Q^k_{\epsilon}}{kn}\left(2\not n
k^2\,(P_1(k)-P_0(k))+2\,P_2(k)(\not k \,pn-\not p \,kn+\frac{k^2}{2}\not n)\right).
\end{eqnarray}
Then the contribution of topology (e)  
to the parton density  can be written as\footnote{Note that a factor of two
has to be included for diagrams which are not symmetric.}
(see Eq.~(\ref{9a}))
\be
\hat\Gamma_{q/q}^{(1,e)}(x,\ep)=
  PP\big\{\frac{1}{16\pi^2}\,\frac{(4\pi)^{\ep}}{\Gamma(1-\ep)}
\int_0^{Q^2} d (-k^2)(-k^2)^{-1-\ep}\,x\,(1-x)^{-\ep}\cdot 2\,
T^{(e)}(x,k^2,\ep)\big\}\label{Tdef}
\ee
where
\begin{eqnarray}
T^{(e)}(x,k^2,\ep)&=&C_F\,\frac{ig^2}{(k^2)^2}\,Tr\,[\frac{\not n}{4kn}\not k\Sigma\not k \gamma^{\mu}\not p\gamma^{\nu}\not
k]\,d_{\mu\nu}(p-k)\nonumber\\
&=&-C_F^2\,\alpha_S^2\,(4\pi)^{\epsilon}\Gamma(1+\epsilon)\,(-k^2)^{-\ep}\nonumber\\
&&
\cdot
\frac{2}{x}\,\Big\{\hat
P_{q/q}(x,\epsilon)\,[T_0\,(\epsilon-1)+4\,(P_1(k)-P_0(k))+\frac{2}{x}\,P_2(k)]\nonumber\\
&&+2P_2(k)\,\frac{(1+x)}{1-x}\Big\}\,.\label{tr}
\end{eqnarray}
Details about the phase space integral are given in Appendix B.\\
$\hat P_{q/q}(x,\ep)$ is the leading order splitting function
in $m=4-2\ep$ dimensions
\be
\hat P_{q/q}(x,\ep)=\frac{1+x^2}{1-x}-\ep\,(1-x).
\ee

\subsubsection{Vertex correction}

The contribution from diagram (d) can also be split into
a Feynman part and an axial part
\be
\Gamma_{\mu,b} = \Gamma_{\mu,b}^F +\Gamma_{\mu,b}^A
\ee
which can written in terms of form factors of
 simple scalar, vector and
tensor loop integrals 
 given in Appendix A. Inserting this expression into
Eq.~(\ref{9a}) and using a definition for $T^{(d)}$
similar to the one we used for diagram (e)
we obtain for the Feynman part
\begin{eqnarray}
T^{(d_F)}(x,k^2,\ep)&=&-C_F(C_F-\frac{N_c}{2})\,
\alpha_S^2\,(4\pi)^{\epsilon}\Gamma(1+\epsilon)\,(-k^2)^{-\ep}\nonumber\\
&&\cdot\frac{4}{x}\,\Big\{\big\{R_1-R_0+(2-\epsilon)R_2-(1-\epsilon)(R_4+R_5)\big\}\,\big[x+\epsilon\,(1-x)\big]\nonumber\\
&&+2\,R_6\,(1-\epsilon)^2\,\hat P_{q/q}(x,\epsilon)\Big\}\label{pretf2}
\end{eqnarray}
and for the axial part
\begin{eqnarray}
T^{(d_A)}(x,k^2,\ep)
&=&-C_F(C_F-\frac{N_c}{2})\,\alpha_S^2\,(4\pi)^{\epsilon}\Gamma(1+\epsilon)\,(-k^2)^{-\ep}\nonumber\\
&&\cdot\Bigg\{\frac{4}{x}\,\left((P_0(k)-P_1(k))\,
\hat P_{q/q}(x,\epsilon)-P_2(k)\,\frac{1+x}{1-x}\right)\nonumber\\
&&+(p^2/k^2)^{-\epsilon}\,\frac{4}{x}\,(P_0(p)-P_1(p)-P_2(p))\,
\hat P_{q/q}(x,\epsilon)\nonumber\\
&&+\frac{2}{x}\,\Big[(S_1+S_4+(2-x)\,S_2-2\,S_0+R_0)\,\hat P_{qq}(x,\epsilon)\nonumber\\
&&-2(R_1+R_2-R_0)\,\frac{1+x}{1-x}\Big]\Bigg\}.\label{pretra2}
\end{eqnarray}
The form factors $R_i,P_j,S_l$; $i=0,\ldots,6$; $j=0,\ldots,2$; $l=0,\ldots 4$
are defined and given in Appendix A both for PV and ML
prescription.

\subsubsection{Ultraviolet renormalization}
We use dimensional regularization to regulate both the
ultraviolet and infrared singularities. Since we do first the
loop integrals we must assume that in $m-4\ep$, $\ep$ is positive.
After adding the counter terms for the quark selfenergy and  vertex
one-loop subdiagrams 
we obtain an ultraviolet finite answer. Then we can analytically
continue the result  to negative values of $\ep$ and we can
go on with the evaluation of the Feynman parameter integrals.
First  we should keep $p$  off-shell
and set it  on-shell only after ultraviolet renormalization.
It is known that the sum of the two ultraviolet counter terms for the vertex
and quark selfenergy diagrams
in regular gauges vanishes as a consequence of abelian gauge invariance.
Keeping only the ultraviolet pole part of the form factors 
$T_0,R_6,P_i$ (see Appendix A) and using Eqs.~(\ref{tr},\ref{pretf2},\ref{pretra2})
one can see that the sum of the ultraviolet contributions
is zero in the ML case, but  is non-vanishing in the PV case.
In particular one finds
\bea
T^{(e),ML}_{UV}&=&C_F^2\as^2\frac{1}{\ep_{UV}}\,\frac{2}{x}\,[-3\,p_{qq}(x)+2]\nonumber\\
T^{(d),ML}_{UV}&=&C_F(C_F-\frac{N_c}{2})\,\as^2\frac{1}{\ep_{UV}}\,\frac{2}{x}\,[3\,p_{qq}(x)-2]\nonumber\\
T^{(e),PV}_{UV}&=&C_F^2\as^2\frac{1}{\ep_{UV}}\,\frac{2}{x}\,p_{qq}(x)\,[-3+4I_0+4\ln{x}]\nonumber\\
T^{(d),PV}_{UV}&=&C_F(C_F-\frac{N_c}{2})\,\as^2\frac{1}{\ep_{UV}}\,\frac{2}{x}\,p_{qq}(x)\,[3-4I_0-2\ln{x}]\nonumber\\
p_{qq}(x)&=&\frac{1+x^2}{1-x} \nonumber\\
\mbox{and so}&&\nonumber\\
T^{(e+d),{\rm ML}}_{\rm UV}&=&0\nonumber\\
T^{(e+d),{\rm PV}}_{\rm UV,C_F^2}&=&C_F^2\as^2\frac{1}{\ep_{UV}}\,\frac{4}{x}\,p_{qq}(x)\,\ln{x}.\label{leftover}
\eea
This is an essential difference between the PV and ML schemes. The leftover UV
singularity (\ref{leftover}) in the PV scheme is a remainder from the
contributions of the spurious poles to the UV renormalization constants. 
In the ML prescription, there are also additional terms which are even not
proportional to the Born term $p_{qq}(x)$, stemming from the $n^*-$part of the
virtual integrals, but those terms completely cancel. Therefore it does not
matter in the ML case whether they are subtracted or not. In the PV case
however, the leftover term in (\ref{leftover}) leads to an additional
contribution to the anomalous dimension in the following way: 
Subtracting all UV poles,
normal and spurious ones, means that in the external loop integral\footnote{We
can consider the last integral over $k^2$ as either infrared or ultraviolet
divergent, depending on where we put a cutoff. CFP treat its divergence as a
collinear  infrared singularity, using an ultraviolet cutoff $Q^2$. Collins
and Soper\,\cite{CSeik} instead introduce an infrared cutoff and consider the last integral
as an ulraviolet divergent integral.} over $k^2$, $T^{(e+d),PV}$ is replaced by
its subtracted value
\be
T^{(e+d),{\rm PV}}_{\rm R}=
T^{(e+d),{\rm PV}}-T^{(e+d),{\rm PV}}_{\rm UV}.\label{tuv}
\ee
 Note that
$T^{(e+d),{\rm PV}}$ 
contains a factor of $(-k^2)^{-\ep}$ whereas $T^{(e+d),{\rm PV}}_{\rm UV}$
does not. Inserting the  UV 
subtracted expression (\ref{tuv}) into the $k^2-$integral (\ref{Tdef}) and
disregarding the finite parts of $T^{(e+d),{\rm PV}}_{\rm R}$ is 
leading to 
\bea
&&PP\{\frac{1}{16\pi^2}\,\frac{(4\pi)^{\ep}}{\Gamma(1-\ep)}
\int^{Q^2}_{0} d(-k^2)(-k^2)^{-1-\ep}\,x\,(1-x)^{-\ep}\cdot 2\,T^{(e+d),{\rm PV},sing}_{\rm R}\}\nonumber\\
&=&C_F^2 (\frac{\as}{2\pi})^2(1-x)^{-\ep}\int^{Q^2}_{0}
d(-k^2)(-k^2)^{-1-\ep}\left[
 (-k^2)^{-\ep}\hat P_{q/q}(x,\ep)\,2\,\frac{\ln x}{\ep}
- p_{qq}(x)\,2\,\frac{\ln x}{\ep}\right]
\nonumber\\&=&\frac{1}{2}\,C_F^2 (\frac{\as}{2\pi})^2
\left(\frac{2\,\ln x}{\ep^2} \,p_{qq}(x)
+\frac{2\,\ln{x}}{\ep}\,[1-x+p_{qq}(x)\,\ln{(1-x)}] + {\cal
O}(\ep^0)\right). 
\eea
The double pole term  has to be dropped since
$\hat\Gamma_{q/q}(x,\alpha_S,\epsilon)$ is defined through simple poles
in $\ep$ (see Eq.~(\ref{gamahat})). But the single pole term gives a
contribution to $\hat\Gamma_{q/q}(x,\alpha_S,\epsilon)$ which is exactly
the difference of the virtual contributions (PV--ML)$_{virt}$ obtained by
using PV
respectively ML 
prescription, as can be seen in Tables \ref{virt} and \ref{rv}. This difference will be
compensated by the ghost diagrams present only in the ML case, as will be
explained in Section \ref{realsec}. \\ 
The complete contributions
of the  virtual diagrams
to $\hat\Gamma^{(1)}_{q/q}(x,\ep)$ are listed in Table ~\ref{virt}. Virtual
diagrams where the cut line is a ghost line would give a contribution at $x=1$
only, so they do not contribute to  $\hat\Gamma^{(1)}_{q/q}(x,\ep)$ (see
Eqs. ~(\ref{gamahat}),~(\ref{hatp})). 
%%%%%%%%%%%%%%%%%%%%%%%%%%%%%%%%%%%%%%%%%%%%%%%%%%%%%%%%%%%%%
\begin{table}
\begin{center}
\begin{tabular}{|l||c|c|c|c|c|c|}
\hline
\multicolumn{7}{|c|}{Finite part}\\
\hline
&$e_{ML}$&$d_{ML}$&$e_{PV}$&$d_{PV}$&$(e+d)_{ML}$&$(e+d)_{PV}$\\
\hline
&&&&&&\\
$p_{qq}(x)$&6&-6&7&-7&0&0\\
$p_{qq}(x)\,\ln^2{x}$&0&0&-2&2&0&0\\
$p_{qq}(x)\,\ln{x}\ln{(1-x)}$&0&0&-4&2&0&-2\\
$p_{qq}(x)\,\ln{x}$ &-1&1&0&0&0&0\\
$p_{qq}(x)\,\ln{(1-x)}$&3&-3&3&-3&0&0\\
$p_{qq}(x)\,Li_2(1-x)$&4&-2&0&2&2&2\\
$p_{qq}(x)\,\pi^2/3$&-2&2&-2&2&0&0\\
$x/(1-x)$&2&-2&0&0&0&0\\
$x$&0&1&-3&4&1&1\\
$1$&-4&4&3&-3&0&0\\
$x\,\ln{x}$&-1&1&4&-2&0&2\\
$\ln{x}$&1&-1&-4&2&0&-2\\
$\ln{(1-x)}$&-2&2&0&0&0&0\\
\hline
\multicolumn{7}{|c|}{Spurious poles}\\
\hline
$p_{qq}(x)\,I_0\,\ln{(1-x)}$&0&0&-4&4&0&0\\
$p_{qq}(x)\,I_0\,\ln{x}$&0&0&-4&4&0&0\\
$p_{qq}(x)\,I_0$ &0&0&0&0&0&0\\
$p_{qq}(x)\,I_1$ &0&0&4&-4&0&0\\
$x\,I_0$&0&0&4&-4&0&0\\
$I_0$&0&0&-4&4&0&0\\
\hline
\multicolumn{7}{|c|}{Singular part}\\
\hline
$p_{qq}(x)\,I_0/\epsilon_{UV}$&0&0&4&-4&0&0\\
$p_{qq}(x)\,\ln{x}/\epsilon_{UV}$&0&0&4&-2&0&2\\
$p_{qq}(x)/\epsilon_{UV}$&-3&3&-3&3&0&0\\
$1/\epsilon_{UV}$&2&-2&0&0&0&0\\
\hline
\end{tabular}
\end{center}
\caption{The results for the virtual
 diagrams after UV subtraction. The virtual contribution to
$\hat\Gamma^{(1)}_{q/q}(x,\ep)$ is obtained as a sum of the finite entries of this
table multiplied by
$-\frac{1}{2\ep}\,C_F^2\,(\frac{\as}{2\pi})^2$. The singular part shows the
ultraviolet counter terms.}\label{virt} 
\end{table}

\subsection{Real contributions}\label{realsec}
In this section we give some  details concerning the
 calculation of  the contributions from the diagrams of topologies (b),(c),(h)
and (i). 
Topology (i) represents the subtraction terms 
${\cal P}(K_0{\cal  P}K_0)$
in Eq.~(\ref{K0exp}) and consists of two Born diagrams linked by an additional
projection. 

\subsubsection{Standard contribution}
The typical integral we have to evaluate for topologies (b),(c) and (h) can be
written as 
\begin{equation}
I^{(a)}=\frac{1}{2}\frac{g^4}{(2\pi)^m}\int \frac{dk^2}{k^4}\,d^{(m-2)}\vec{k}\,\int
d\Phi(l_1,l_2)\cdot M^{(a)}(x,l_1,l_2,k,\ep)
\label{ij}
\end{equation}
where $d\Phi(l_1,l_2)$ is the two-body-phase space for the two cut
lines and $M^{(a)}$ the matrix element for topology (a). 
In terms of light-cone variables $q_{\mu}$ $=$ $(q^+,q^-,\vec{q})$
\bea
d\Phi(l_1,l_2) &=&\frac{4\pi^2}{(2\pi)^m}\,
 \int dl_1^+ dl_2^+ dl_1^- dl_2^-
\,d^{(m-2)}
\vec{l}_1 d^{(m-2)}\vec{l}_2\,\delta(p^+-k^+-l_1^{+}-l_2^{+})\nonumber\\
&&\cdot\delta^{(m-2)}(\vec{k}+\vec{l}_1+\vec{l}_2)\,
\delta(k^-+l_1^-+l_2^-)\,\delta(l_1^2)\,\delta(l_2^2)\,.
\eea
Introducing dimensionless parameters $z_j=l^+_j/p^+$, $j=1,2$
and using the relation \\ $q^-=(q^2+\vec{q}^2)/(2q^+)$ on gets
\bea
d\Phi(l_1,l_2) &=&\frac{2\pi^2}{(2\pi)^m}\, \int \frac{dz_1}{z_1}\frac{dz_2}{z_2}\,d\vec{l}_1d\vec{l}_2\,\delta(1-x-z_1-z_2)\nonumber\\
&&\cdot\delta\,(\vec{k}+\vec{l}_1+\vec{l}_2)\,\delta(\frac{k^2}{x}+\frac{\vec{k}^2}{x}+\frac{\vec{l}_1^2}{z_1}+\frac{\vec{l}_2^2}{z_2}).
\nonumber
\eea
The external loop also requires  the integration over $d\vec{k}$.
The transverse momentum integrals can be carried out easily
 if we introduce new momenta such
that i) the $\delta-$function will be diagonal,
ii) the propagator denominators appearing in the matrix element
become as simple as possible.
The appropriate choice is different for different topologies.
For example, for toplogy (b) and (h) the good variables are
$$
\vec{l}_1=\vec{h}_1 \hspace*{1cm}
\vec{l}_2=-\frac{z_2}{1-z_1}(\vec{h}_1+\vec{h}_2)\hspace*{1cm}\vec{h}_1\vec{h}_2= h_1h_2\cos{\theta}\,.
$$
For the  phase space integral
we  obtain  the form
\bea
PS&=&\frac{2\pi^{4-2\epsilon}}{(2\pi)^{m}}\,\frac{1}{\Gamma^2(1-\epsilon)}
\,(-k^2)^{1-2\epsilon}\int_0^{1-x} dz_1 \,
\left(\frac{z_1(1-x-z_1)}{x}\right)^{-\epsilon}\nonumber\\
&& \cdot\int_0^{1} du\,
u^{-\epsilon}\,(1-u)^{-\epsilon}\cdot\frac{\Gamma(1-\epsilon)}{\sqrt{\pi}\Gamma(\frac{1}{2}-\epsilon)}\int_0^{\pi}d\theta\,(\sin\theta)^{-2\epsilon}\label{psb}
\eea
where $PS$ is defined as $PS=\int d\vec{k}\,d\Phi(l_1,l_2)$.
The azimuthal dependence of the denominator is either trivial or it can be cast
into the form $1/(1+\lambda^2+2\lambda\cos{\theta})$, such that we need the
integrals 
\begin{eqnarray}
I_A&=&\frac{\Gamma(1-\epsilon)}{\sqrt{\pi}\Gamma(\frac{1}{2}-\epsilon)}\int_0^{\pi}d\theta\,(\sin\theta)^{-2\epsilon}=1\nonumber\\
I_C(\lambda^2)&=&\frac{\Gamma(1-\epsilon)}{\sqrt{\pi}\Gamma(\frac{1}{2}-\epsilon)}\int_0^{\pi}d\theta\,\frac{(\sin\theta)^{-2\epsilon}}{1+\lambda^2+2\lambda\cos{\theta}}
\nonumber\\
&&\nonumber\\
I_C(\lambda^2)&=&\left\{\begin{array}{ll}
F(1,1+\epsilon;1-\epsilon,\lambda^2)\qquad\mbox{ for }\lambda^2 <1\\
\frac{1}{\lambda^2}\,F(1,1+\epsilon;1-\epsilon,\frac{1}{\lambda^2})\quad\mbox{ for }\lambda^2 >1.
\end{array}\right.\label{ic}
\end{eqnarray}
In the case of topology (b)  for example one has
\bea
\lambda^2&=&\frac{b\,u}{1-u}\nonumber\\
b&=&\frac{z_1(1-x-z_1)}{x}\nonumber
\eea
such that the integration range for the $u-$integration will be split at $u
=\frac{1}{1+b}$ according to Eq.~(\ref{ic}). 

Carrying out these integrals one should keep the regularization
parameter of the PV scheme appearing in Eq.~(\ref{pv})
to regularize the spurious poles. As we explained above, the contibutions from 
diagrams $(c)$, $(b^{D_{11}})$, $(h^{D_{11}})$ and $(i^{D_{11}})$
are the same for the PV and the ML scheme. There are no ghosts related to
diagram (c) since it does not contain cut gluon lines. 
The individual contributions are shown in Tables \ref{qq} and \ref{qqsi}.
The new feature of the ML scheme is that we should also
add the contributions from the axial ghost terms (see Eq.~(\ref{disprop}) and Fig.~2).
%%%%%%%%%%%%%%%%%%%%%%%%%%%%%%%%%%%%%%%%%%%%%%%%%%%%%%%%%%%%
\begin{table}
\lspage{
\begin{center}
\begin{tabular}{|l||c|c|c|c||c|c|c|c|c|c||c|c|}
\hline
&\multicolumn{4}{|c||}{no ghosts}&\multicolumn{6}{|c||}{ghosts}&no ghosts&ghosts\\
\hline
&$b^{D_{11}}$&$(c)$&$h^{D_{11}}$&$i^{D_{11}}$&$b^{D_{12}}$&$b^{D_{21}}$&$h^{D_{12}}$&$h^{D_{21}}$&$i^{D_{12}}$&$i^{D_{21}}$&$(b+h-i)_{11}$&$(b+h-i)_{gh}$\\
\hline
&&&&&&&&&&&&\\
$p_{qq}(x)\,\ln^2{(1-x)}$&0&0&6&6&0&0&0&0&0&0&0&0\\
$p_{qq}(x)\,\ln^2{x}$&-1&-1&0&-2&1/2&1/2&1&0&2&0&1&0\\
$p_{qq}(x)\,\ln{x}\ln{(1-x)}$&0&0&-2&-2&0&0&-2&0&0&0&0&-2\\
$p_{qq}(x)\,\ln{(1-x)}$&4&0&-4&0&&&&&&&0&0\\
$p_{qq}(x)\,\ln{x}$&0&-3/2&0&0&&&&&&&0&0\\
$p_{qq}(x)\,Li_2(1-x)$&0&-2&0&0&1&1&-2&0&0&0&0&0\\
$p_{qq}(x)\,\pi^2/3$&0&0&-2&-2&&&&&&&0&0\\
$x\,\ln^2{x}$&0&0&-1/2&0&&&&&&&-1/2&0\\
$\ln^2{x}$&0&0&-1/2&0&&&&&&&-1/2&0\\
$x\,\ln{x}$&2&-7/2&-4&0&0&0&2&0&0&0&-2&2\\
$\ln{x}$&2&-7/2&0&-2&0&0&-2&0&0&0&4&-2\\
$(1+x)Li_2(1-x)$&0&0&2&2&&&&&&&0&0\\
$1-x$&0&-7&1&-2&&&&&&&3&0\\
1&0&-1&0&0&&&&&&&0&0\\
$x^2\,\ln{x}\ln{(1-x)}/(1-x)$&0&0&-4&-4&&&&&&&0&0\\
\hline
\multicolumn{13}{|c|}{Spurious poles}\\
\hline
$p_{qq}(x)\,I_0$&4&0&0&4&-2&-2&0&0&0&-4&0&0\\
$p_{qq}(x)\,I_0\,\ln{x}$&0&0&-4&-4&0&0&2&2&4&0&0&0\\
$p_{qq}(x)\,I_0\,\ln{(1-x)}$&0&0&4&4&0&0&-2&-2&0&-4&0&0\\
$p_{qq}(x)\,I_1$&0&0&4&4&0&0&-2&-2&-4&0&0&0\\
$I_0\cdot x/(1-x)$&0&0&-8&-8&0&0&4&4&0&8&0&0\\
\hline
\end{tabular}
\end{center}
\caption{Results for real diagrams. The contribution to
$\hat\Gamma^{(1)}_{q/q}(x,\ep)$ is obtained as a sum of the entries 
multiplied by  $-\frac{1}{2\ep}\,C_F^2\,(\frac{\as}{2\pi})^2$}\label{qq}
}
\end{table}

\begin{table}
\lspage{
\begin{center}
\begin{tabular}{|l||c|c|c|c||c|c|c|c|c|c||c|c|}
\hline
&\multicolumn{4}{|c||}{no ghosts}&\multicolumn{6}{|c||}{ghosts}&no ghosts&ghosts\\
\hline
&$b^{D_{11}}$&$(c)$&$h^{D_{11}}$&$i^{D_{11}}$&$b^{D_{12}}$&$b^{D_{21}}$&$h^{D_{12}}$&$h^{D_{21}}$&$i^{D_{12}}$&$i^{D_{21}}$&$(b+h-i)_{11}$&$(b+h-i)_{gh}$\\
\hline
&&&&&&&&&&&&\\
$p_{qq}(x)\,I_0/\epsilon$&0&0&-4&-4&0&0&2&2&2&2&0&0\\
$p_{qq}(x)\,\ln{x}/\epsilon$&0&0&0&0&0&0&2&0&2&0&0&0\\
$p_{qq}(x)\,\ln{(1-x)}/\epsilon$&0&0&-4&-4&0&0&0&0&0&0&0&0\\
$(1+x)\,\ln{x}/\epsilon$&0&0&-1&-1&0&0&0&0&0&0&0&0\\
$(1-x)/\epsilon$&0&0&2&2&0&0&0&0&0&0&0&0\\
&&&&&&&&&&&&\\
\hline
\end{tabular}
\end{center}
\caption{Real diagrams, singular parts}\label{qqsi}
}
\end{table}

\begin{table}[hp!]
\lspage{
\begin{center}
\begin{tabular}{|l||c|c|c||c|c|c|c|}
\hline
&\multicolumn{2}{|c|}{real sum}&diff$_{real}$&\multicolumn{2}{|c|}{virtual
sum}&diff$_{virt}$&\\
\hline
&&&&&&&\\
&$(b+h-i)_{ML}$&$(b+h-i)_{PV}$&(ML--PV)$_{real}$&$(e+d)_{ML}$&$(e+d)_{PV}$&(ML--PV)$_{virt}$&sum of ghosts\\
&&&&&&&only\\
\hline
&&&&&&&\\
$p_{qq}(x)\,\ln^2{x}$&1&1&0&0&0&0&0\\
$p_{qq}(x)\,\ln{x}\ln{(1-x)}$&-2&0&-2&0&-2&2&-2\\
$p_{qq}(x)\,Li_2(1-x)$&0&0&0&2&2&0&0\\
$x\,\ln^2{x}$&-1/2&-1/2&0&0&0&0&0\\
$\ln^2{x}$&-1/2&-1/2&0&0&0&0&0\\
$x\,\ln{x}$&0&-2&2&0&2&-2&2\\
$\ln{x}$&2&4&-2&0&-2&2&-2\\
1&3&3&0&0&0&0&0\\
$x$&-3&-3&0&1&1&0&0\\
&&&&&&&\\
\hline
\end{tabular}
\end{center}
\caption{Comparison of real and virtual parts}\label{rv}
}
\end{table}

\subsubsection{Ghost contributions}

The diagrams considered in this section are given by topologies $b^{D_{12}}$,
$h^{D_{12}}$, $h^{D_{21}}$, $i^{D_{12}}$ and $i^{D_{21}}$. Topology $b^{D_{21}}$ leads to the same result
as $b^{D_{12}}$ because the diagrams are symmetric under exchange of
$l_1\leftrightarrow l_2$. Diagrams with two cut ghost lines only give a
contribution at $x=1$, which has been omitted since it can be obtained more
easily from fermion number conservation (see Section \ref{fn}).

\medskip

The phase space for diagrams with one ghost line and one gluon line cut is
given by (for details see Appendix  \ref{PD12})
\begin{eqnarray}
PS^{ghost}&=&F_{\epsilon}\,|k^2|^{2-2\epsilon}\,x^{-1+\epsilon}(1-x)^{-\epsilon}\int_0^1 du\,
u^{-\epsilon}(1-u)^{1-\epsilon}\int_0^1 dy\,y^{-\epsilon}\nonumber\\
&&\frac{1}{B(\frac{1}{2}-\epsilon,\frac{1}{2}-\epsilon)}\int_0^1
dw\,[w(1-w)]^{-\frac{1}{2}-\epsilon}. \label{psd12}
\end{eqnarray}
The results  for the ghost diagrams are given in  Tables \ref{qq} and
\ref{qqsi}. Table \ref{rv} shows the results for both, the real and the
virtual part in the PV and ML schemes, where real means standard {\it plus}
ghost contributions in the ML case.

\section{Discussion and Conclusions}

The reevaluation of the two-loop anomalous dimension
in light-like axial gauge is a good testing ground to understand
the usefulness and reliability of this gauge.
 The success of the
calculation of Curci, Furmanski and Petronzio using  the principal value
prescription for the $1/nq$ factor is based on some heuristic
treatment of the ultraviolet renormalization. Since the anomalous
dimensions are related to the ultraviolet renormalization of bilocal
operators, it is important to provide a better field theoretical support
for the "phenomenologica rules'' found by CFP. 
Within the PV scheme one gets spurious UV singular terms
which are subtracted.
The Mandelstam-Leibbrandt prescription eliminates this
difficulty and  is in addition
consistent with canonical quantization. Therefore  
it has to be expected  that the
straightforward  use of the Feynman rules as given
by the ML scheme naturally provides
a field theoretically correct treatment, leading to a deeper understanding
of the heuristic CFP  rules. This expectation 
has been justified by our explicit calculation of the
 two-loop anomalous dimensions  proportional to the colour factor 
$C_F^2$. 

It is instructive to make a detailed comparison  of the
 evaluations in the PV and ML schemes. The necessary informations are
summarized in Tables 1,~2,~3 and 4. In evaluating one-loop insertions in
the ML-scheme, the spurious poles cancel within the loop integrals, 
such that the ML result is free from spurious singularities, 
whereas the PV integrals can be recovered
as a subpart of the ML integrals, this subpart being plagued by  spurious
singularities (see Eq.~(\ref{virtPV}) and Table \ref{virt}). 
After having carried out the momentum integration with ML prescription, one
can regularize the spurious poles of Feynman-parameter integrals with a
PV prescription. 
If   all singularities
are regularized 
with dimensional regularization the answer for the sum remains the same.
The sums of all virtual
 contributions are different in the two schemes and the difference
is due to the difference in the ultraviolet subtraction.

The differences in the real contributions can be organized according to
 Eq.~(\ref{disprop}). The first term in this equation is the standard
axial gauge contribution.
The second term defines the axial ghost contribution.
The results are summarized in Tables \ref{qq},~\ref{qqsi} and \ref{rv}.
The sum of all real contributions is different
in the two schemes, the  difference being given by the axial ghost
contribution which
compensates the difference found in  the
virtual corrections.
We see again that the individual terms in the ML scheme are more
regular. If we combine the contributions of the diagrams 
$(\,b^{D_{11}}$,$~b^{D_{12}}$,$~b^{D_{21}})$ or
$(\,h^{D_{11}}$,$~h^{D_{12}}$,$~h^{D_{21}})$ or
$(\,i^{D_{11}}$,$~i^{D_{12}}$,$~i^{D_{21}})$ their sum
is separately finite. In addition, the sums of the ghost and non-ghost
diagrams are also separately finite. 
The fact that the spurious singularities cancel separately in the
sum of the virtual and  real contributions is a consequence of
gauge invariance. 
  
In summary, the evaluation of the two-loop anomalous dimension
in the ML scheme is a
consistent  method. The axial ghost contributions are important
to get the correct answer. It is remarkable that the
phenomenological rule of CFP for subtracting all ultraviolet
contributions
(spurious and non-spurious ones) leads to the same additional terms
as provided by the axial ghost contributions in the ML scheme.\\
It would be interesting to see the differences between the
two schemes and the higher
consistency
of the ML scheme also for the remaining colour structures.

\newpage

\appendix

\section{Virtual integrals}
We define general
n-point integrals, containing no axial denominator $1/qn$, by
\begin{displaymath}
J^{F\mu_1\ldots\mu_s}_n(k_1\ldots k_{n-1})= \int d^m q \frac{q^{\mu_1}\ldots q^{\mu_s}}{q^2\,(q+k_1)^2\ldots (q+k_{n-1})^2}
\end{displaymath}
and   n-point integrals containing one axial denominator $1/qn$ by
\begin{eqnarray*}
J^{A\mu_1\ldots\mu_s}_n(k_1\ldots k_{n-1})&=&\int d^m q \frac{q^{\mu_1}\ldots
q^{\mu_s}}{q^2\,(q+k_1)^2\ldots (q+k_{n-1})^2\,qn}\\
&&\\
I^{A\mu_1\ldots\mu_s}_n(k_1\ldots k_{n})&=&\int d^m q \frac{q^{\mu_1}\ldots
q^{\mu_s}}{(q+k_1)^2\ldots (q+k_{n})^2\,qn}
\end{eqnarray*}
In the calculation of the contribution of the diagrams of type (e) and
(d)
we need 
two-point integrals and three-point integrals.
\subsection{Two-point integrals}
First we introduce a general parametrization of the integrals 
as follows
\begin{eqnarray*}
J_2^F(r) &=& Q^r_{\epsilon}\, T_0\qquad;\qquad T_0=\frac{1}{\epsilon_{UV}}+2\\
J_2^{F\mu}(r)&=&-Q^r_{\epsilon}\,T_1  r^{\mu}\qquad;\qquad
T_1=\frac{1}{2}\,T_0\\
&&\\
J_2^A(r) &=& -\frac{Q^r_{\epsilon}}{rn}\,P_0(r)\\
J_{2\mu}^{A}(r)&=&\frac{Q^r_{\epsilon}}{rn}\,\{P_1(r)\, r_{\mu} + P_2(r)\,
n_{\mu}^{*} +P_3(r)\,n_{\mu}\}\\
&&\\
Q^r_{\epsilon} &=& i\pi^{\frac{m}{2}}\Gamma (1+\epsilon)(-r^2)^{-\epsilon}.
\end{eqnarray*}
The results for the form factors $P_0, P_1, P_2$ and $P_3$ 
depend on the regularization of the axial denominator $1/qn$.

\subsubsection{Form factors for the ML scheme}
If we use ML prescription we get
\begin{eqnarray}
P_0^{ML}(r)&=&Li_2(1)-Li_2(1-\chi_r)\label{2genML}\\
P_1^{ML}(r)&=&-\frac{\chi_r\ln{\chi_r}}{1-\chi_r}\\
P_2^{ML}(r)&=&\frac{rn}{nn^*}\,\left(\frac{1}{\epsilon_{UV}}+2+\frac{\chi_r\ln{\chi_r}}{1-\chi_r}\right)\\
P_3^{ML}(r)&=&\frac{rn^*}{nn^*}\,\left(-1+\frac{\ln{\chi_r}}{1-\chi_r}+\frac{1}{\chi_r}\,[Li_2(1)-Li_2(1-\chi_r)]\right)
\end{eqnarray}
where $\chi_r$ is defined as
\begin{displaymath}
\chi_r=:\frac{2\,rn\,rn^*}{nn^*\,r^2}
\end{displaymath}

\subsubsection{Form factors for the PV scheme}
In the case of PV prescription
on gets
\begin{eqnarray}
P_0^{PV}(r) &=&\frac{I_0}{\epsilon_{UV}}+\frac{\ln{(r^+)}}{\epsilon_{UV}}-I_1+
I_0\ln{(r^+)}+\frac{1}{2}\ln^2{(r^+)}+ Li_2(1)\\
P_1^{PV}(r)&=&\frac{1}{\epsilon_{UV}}+2\\
P_2^{PV}(r)&=&0\\
P_3^{PV}(r)&=&\frac{r^2}{2\,rn}\left\{\frac{I_0}{\epsilon_{UV}}+\frac{\ln{(r^+)}}{\epsilon_{UV}}-\frac{2}{\epsilon_{UV}}-I_1+I_0\ln{(r^+)}+\frac{1}{2}\ln^2{(r^+)}-4+Li_2(1)\right\}.\nonumber\\
&&\hfill\label{2genPV}
\end{eqnarray}
There is no $n^*-$dependence  in this case and we used the definitions
\begin{eqnarray*}
r^+ &=&\frac{rn}{pn}\\
&&\\
I_0 &=&\int_0^1du\,\frac{u}{u^2+\delta^2}= -\ln{\delta}\,+\,{\cal O}(\delta)\\
I_1 &=&\int_0^1du\,\frac{u\,\ln{u}}{u^2+\delta^2}=
-\frac{1}{2}\,\ln^2{\delta}-\frac{1}{4}\,Li_2(1)\,+\,{\cal O}(\delta)\,.
\end{eqnarray*}

\subsection{Two-point integrals for special momenta $k$ and $p$}
The parameters for the two-point integrals with the special kinematics used
in the calculation can be read off from relations (\ref{2genML}) to
(\ref{2genPV}). In this case 
\begin{eqnarray}
p&=&n^*\label{nst}\\
(p-k)^2&=&0\quad\Rightarrow \quad pk = \frac{k^2}{2}\quad;\quad k_{T}^2 =
-k^2(1-x)\quad;\quad
k^2<0\\
k^+&=&x\quad;\quad p^+=1\label{ppl}\\
\chi_k&=&\frac{2\,kn\,kn^*}{nn^*\,k^2}=x\nonumber\\
\chi_p&=&\frac{2\,p^+\,p^-}{\,p^2}=1\quad\mbox{for }\,p^2\not =0, \,p_{T}^2=0. \nonumber
\end{eqnarray}

Note also that $$\lim_{\chi_p\to 1}\frac{\chi_p\ln{\chi_p}}{1-\chi_p}=-1~.$$\\
Since the integrals $J_2^A(p)$ and $J_{2\mu}^A(p)$ only contribute to the UV
counterterm, $p^2$ has to be off-shell in this case. In the infrared region,
where $\epsilon <0$ and $p^2=0$, the integrals $J_2^A(p)$ and $J_{2\mu}^A(p)$
vanish due to the overall factor $(-p^2)^{-\epsilon}$.

\subsubsection{Form factors in the ML scheme}

\begin{eqnarray*}
P_0^{ML}(k)&=&Li_2(1)-Li_2(1-x)\\
P_1^{ML}(k)&=&-\frac{x\ln{x}}{1-x}\\
P_2^{ML}(k)&=&x\,\left\{\frac{1}{\epsilon_{UV}}+2+\frac{x\ln{x}}{1-x}\right\}\\
P_3^{ML}(k)&=&\frac{k^2}{2pn}\,\left\{-1+\frac{\ln{x}}{1-x}+\frac{1}{x}\,[Li_2(1)-Li_2(1-x)]\right\}\\
&&\\
P_0^{ML}(p)&=&Li_2(1)\\
P_1^{ML}(p)&=&1\\
P_2^{ML}(p)&=&\frac{1}{\epsilon_{UV}}+1\\
P_3^{ML}(p)&=&\frac{p^2}{2pn}\,\{-2+Li_2(1)\}
\end{eqnarray*}

\subsubsection{Form factors in the PV scheme}

\begin{eqnarray*}
P_0^{PV}(k) &=&\frac{I_0}{\epsilon_{UV}}+\frac{\ln{x}}{\epsilon_{UV}}-I_1+
I_0\ln{x}+\frac{1}{2}\ln^2{x}+ Li_2(1)\\
P_1^{PV}(k)&=&\frac{1}{\epsilon_{UV}}+2\\
P_2^{PV}(k)&=&0\\
P_3^{PV}(k)&=&\frac{k^2}{2kn}\,\left\{\frac{I_0}{\epsilon_{UV}}+\frac{\ln{x}}{\epsilon_{UV}}-\frac{2}{\epsilon_{UV}}-I_1+I_0\ln{x}+\frac{1}{2}\ln^2{x}-4+Li_2(1)\right\}\\
&&\\
P_0^{PV}(p) &=&\frac{I_0}{\epsilon_{UV}}-I_1+ Li_2(1)\\
P_1^{PV}(p)&=&\frac{1}{\epsilon_{UV}}+2\\
P_2^{PV}(p)&=&0\\
P_3^{PV}(p)&=&\frac{p^2}{2pn}\,\left\{\frac{I_0}{\epsilon_{UV}}-\frac{2}{\epsilon_{UV}}-I_1-4+Li_2(1)\right\}
\end{eqnarray*}

\subsection{Three-point integrals}

The three-point integrals will only be given for the special kinematics needed
in the calculation, defined by the  relations (\ref{nst}) to (\ref{ppl}), since
a general form depending on parameters like $\chi_r$ cannot be so neatly
obtained as it was the case for the two-point integrals. Below we list the
integrals in terms of form factors. 

\begin{eqnarray*}
J_3^F(k,p) &=& -Q^k_{\epsilon}(-k^2)^{-1} R_0\\
J_3^{F\mu}(k,p)&=& Q^k_{\epsilon}(-k^2)^{-1}(R_1 p^{\mu} + R_2  k^{\mu})\\
J_3^{F\mu\nu}(k,p) &=& - Q^k_{\epsilon}(-k^2)^{-1} (R_3 p^{\mu}p^{\nu} + R_4
k^{\mu}k^{\nu}+ R_5\{kp\}^{\mu\nu} + k^2 R_6 g^{\mu\nu})\\
\{kp\}^{\mu\nu}&=&k^{\mu}p^{\nu}+k^{\nu}p^{\mu}\\
J_3^A(k,p) &=&\frac{Q^k_{\epsilon}}{pn}(-k^2)^{-1} S_0\\
J_{3\mu}^{A}(k,p)&=&- \frac{Q^k_{\epsilon}}{pn}(-k^2)^{-1}(S_1 p_{\mu} + S_2
k_{\mu} + S_3 n_{\mu}+ S_4 n^*_{\mu})
\end{eqnarray*}

Higher tensor integrals have been eliminated by
Passarino-Veltman reduction. 

\medskip

\subsubsection{Form factors for the Feynman part}
\begin{center}
\begin{tabular}{|c|c|c|c|c|c|c|}
\hline
$R_0$&$R_1$&$R_2$&$R_3$&$R_4$&$R_5$&$R_6$\\
\hline
&&&&&&\\
$\frac{1}{\epsilon^2}- \frac{\pi^2}{6}$&
$\frac{1}{\epsilon^2}+\frac{2}{\epsilon}+4-\frac{\pi^2}{6}$&
$-\left(\frac{1}{\epsilon}+2\right)$&
$R_1+\frac{1}{\epsilon}+3 $&
$-\left(\frac{1}{2\epsilon}+1\right)$&
$-\left(\frac{1}{2\epsilon}+\frac{3}{2}\right)$&
$\frac{1}{4}\left(\frac{1}{\epsilon_{UV}}+3\right)$\\
&&&&&&\\
\hline
\end{tabular}
\end{center}

\subsubsection{Axial part form factors with ML prescription}

\begin{eqnarray*}
S_0^{ML}&=&\frac{1}{\epsilon^2}+\frac{1}{\epsilon}\,\ln x-2
Li_2(1)-2Li_2(1-x)-\frac{1}{2}\,\ln^2{x} \\
S_1^{ML}&=&\frac{1}{\epsilon^2}+\frac{1}{\epsilon}-\frac{1}{\epsilon}\,\frac{x\ln x}{1-x}-\frac{x\ln x}{1-x}- Li_2(1)+
      \frac{2\,x}{1-x}\,Li_2(1-x)+\frac{1}{2}\,\frac{x}{1-x}\,\ln^2{x}\\
S_2^{ML}&=&\frac{1}{\epsilon}\,\frac{\ln x}{1-x}-
      \frac{2}{1-x}\,Li_2(1-x)-\frac{1}{2}\,\frac{\ln^2{x}}{1-x}\\
S_3^{ML}&=&-\frac{1}{2}\,\frac{k^2}{kn}\,\left(Li_2(1)-\frac{Li_2(1-x)}{1-x}\right)\\
S_4^{ML}&=&-\frac{1}{\epsilon}+\frac{x\ln x}{1-x}
\end{eqnarray*}

\subsubsection{Axial part form factors with PV prescription}

\begin{eqnarray*}
S_0^{PV}&=&\frac{1}{\epsilon^2}-\frac{1}{\epsilon}\,I_0+\frac{1}{\epsilon}\ln{x} +
I_1-I_0\ln{x}-2 Li_2(1) - 2 Li_2(1-x)-\frac{1}{2}\,\ln^2{x}\\
S_1^{PV} &=&  \frac{1}{\epsilon^2}-\frac{1}{\epsilon}\,\frac{x\ln{x}}{1-x} + \frac{x}{1-x}\,Li_2(1-x)- Li_2(1)\\
S_2^{PV} &=& \frac{1}{\epsilon}\,\frac{\ln x}{1-x}-\frac{Li_2(1-x)}{1-x}\\
S_3^{PV}&=&-\frac{1}{2}\,\frac{k^2}{kn}\,\left(\frac{I_0}{\epsilon}+\frac{1}{\epsilon}\frac{\ln
x}{1-x}-I_1+\frac{I_0\ln{x}}{1-x}-Li_2(1)-\frac{xLi_2(1-x)}{1-x}+\frac{1}{2}\,\ln^2{x}\right)\\
S_4^{PV}&=&0
\end{eqnarray*}

\section{Phase space parametrizations}

\subsection{Phase space integral for virtual contributions}

The phase space integral needed for the virtual diagrams, where only one gluon
line is cut, is given by
\begin{equation}
PS^{virt}=2\pi\,z\int\frac{d^mk}{(2\pi)^m}\,\delta(x-z)\,\delta((p-k)^2)
\end{equation}
where 
\begin{displaymath}
(p-k)^2 = -\frac{k_T^2}{x}-\frac{(1-x)}{x}k^2\Longrightarrow
\delta((p-k)^2)=x\,\delta(k_T^2+(1-x)k^2)
\end{displaymath}
Since the integrand has no angular dependence, the angular integral is trivial
here, so 
\begin{displaymath}
\int d^m k =K_{m-2}\,\int dk^2\,\frac{dx}{2x}\,d|k_T|\,|k_T|^{m-3}=K_{m-2}\,\int dk^2\,\frac{dx}{2x}\,\frac{1}{2}\,dk_T^2\left(k_T^2\right)^{\frac{m-4}{2}}
\end{displaymath}
and $$K_{m-2}=\frac{2\,\pi^{\frac{m-2}{2}}}{\Gamma(\frac{m-2}{2})}$$ is the
surface of a $(m-2)$ dimensional hypersphere. Thus in $m=4-2\epsilon$
dimensions, the phase space for the virtual diagrams is given by   
\begin{eqnarray*}
PS^{virt}&=&
2\pi\,z\int \frac{d^m k}{(2\pi)^m}\,\delta(x-z)\,\delta((p-k)^2)\\
&=&\frac{2\pi}{(2\pi)^{4-2\epsilon}}\frac{1}{4}\,K_{2-2\epsilon}\,\int
dk^2\,dk_T^2\left(k_T^2\right)^{-\epsilon}\,x\,\delta(k_T^2+(1-x)k^2)\\
&=&\frac{1}{16\pi^2}\frac{(4\pi)^{\epsilon}}{\Gamma(1-\epsilon)}\int_0^{Q^2}
d(-k^2)\,(-k^2)^{-\epsilon}\,x\,(1-x)^{-\epsilon}
\end{eqnarray*}
The upper limit of the $d(-k^2)$ integral $(k^2<0$) is denoted by a large
momentum scale $Q^2$ whose actual value is irrelevant since only the pole part
of the $k^2-$integration is needed.

\subsection{Phase space integral for  axial ghost contributions}\label{PD12}

The two-body-phase space  where one of the cut lines, say $l_2$,  is an axial
ghost is goverened by the condition $l_2^+ = 0$
\begin{eqnarray}
PS^{ghost}&=&\frac{4\pi^2}{(2\pi)^m}\, \int dl_1^+ dl_2^+ dl_1^- dl_2^-
\,d^{(m-2)}\vec{l}_1 d^{(m-2)}\vec{l}_2\,d^{(m-2)}\vec{k}_T\,\delta(p^+-k^+-l_1^{+}-l_2^{+})\nonumber\\
&&\cdot\delta^{(m-2)}(\vec{k}_T+\vec{l}_1+\vec{l}_2)\,\delta(k^-+l_1^-+l_2^-)\,\delta(l_2^+)\,\delta(l_1^2)\,\theta(l_2^0)\nonumber\\
&&\nonumber\\
\mbox{Now use}\qquad \delta(l_1^2)&=& \frac{1}{2l_1^+}\,\delta(l_1^-
-\frac{\vec{l}_1^2}{2l_1^+})\qquad \mbox{to eliminate }\,l_1^-\nonumber\\
\mbox{substitute } l_2^- \mbox{ by }\xi&=&2p^+l_2^-\Rightarrow
dl_2^-=\frac{d\xi}{2p^+}\nonumber\\
\mbox{Note that }\xi\geq 0&&\mbox{because of the theta function
}\theta(l_2^0):\nonumber\\
\theta(l_2^0)&=&\theta(l_2^++l_2^-)=\theta(l_2^-)\quad\mbox{ for }l_2^+=0\nonumber\\
\mbox{ Hence }&&\nonumber\\
PS^{ghost}&=&\frac{2\pi^2}{(2\pi)^m}\,\int\frac{dz_1}{z_1}dz_2\delta(z_2)\delta(1-x-z_1-z_2)\int
d\vec
k_{T}\,d\vec{l}_1d\vec{l}_2\,\delta\,(\vec{k}_T+\vec{l}_1+\vec{l}_2)\nonumber\\
&&\int_0 d\xi\,\delta(\frac{k^2}{x}+\frac{\vec{k}_T^2}{x}+\xi+\frac{\vec{l}_1^2}{z_1})
\end{eqnarray}
For the diagram $(h^{D_{12}})$, 
it is convenient to eliminate $\vec l_2$ by using the $\delta-$function for
the transverse momenta. 
For diagrams $(b^{D_{12}})$ and $(h^{D_{21}})$ it is more convenient to
eliminate $\vec{k}_T$, leading to a simpler form of the denominators. But the
final form of the phase space of course is the same. So for definiteness, the
substitutions suitable for topology $(h^{D_{12}})$ will be given below.  \\
The angle $\theta$ and the parameter
$\beta$ are defined by 
\begin{eqnarray*}
\vec{k}_T\vec l_1 &=&k_T l_1\cos{\theta}\quad ;\quad 
\beta^2=\frac{k_T^2}{l_1^2}
\end{eqnarray*}
Using
\begin{eqnarray*}
\int d\vec l_1\,d\vec k_T
&=&\frac{\pi^{2-2\epsilon}}{\Gamma^2(1-\epsilon)}\int dl_1^2\,
(l_1^2)^{-\epsilon}\,dk_T^2\, (k_T^2)^{-\epsilon}\cdot\frac{\Gamma(1-\epsilon)}{\sqrt{\pi}\Gamma(\frac{1}{2}-\epsilon)}\int_0^{\pi}d\theta
(\sin\theta)^{-2\epsilon}
\end{eqnarray*}
leads to
\begin{eqnarray}
PS^{ghost}&=&F_{\epsilon}\int\frac{dz_1}{z_1}\delta(1-x-z_1)\int dl_1^2
(l_1^2)^{-\epsilon}dk_T^2 (k_T^2)^{-\epsilon}\nonumber\\
&&\cdot\int_0
d\xi\,\delta(\frac{k^2}{x}+\frac{\vec{k}_T^2}{x}+\xi+\frac{\vec{l}_1^2}{z_1})\cdot\frac{\Gamma(1-\epsilon)}{\sqrt{\pi}\Gamma(\frac{1}{2}-\epsilon)}\int_0^{\pi}d\theta
(\sin\theta)^{-2\epsilon}\\
&&\nonumber\\
F_{\epsilon}&=&\frac{2\pi^{4-2\epsilon}}{(2\pi)^m}\frac{1}{\Gamma^2(1-\epsilon)}.\nonumber
\end{eqnarray}
Now substitute 
\begin{eqnarray*}
k_T^2&=&|k^2|\cdot u\quad;\quad l_1^2=|k^2|\frac{(1-x)}{x}\,(1-u)\cdot y\\
\mbox{ then }&&\\
\beta^2&=&\frac{k_T^2}{l_1^2}=\frac{xu}{(1-x)(1-u)y}\,.
\end{eqnarray*}
This is leading to 
\begin{eqnarray}
PS^{ghost}&=&F_{\epsilon}\,|k^2|^{2-2\epsilon}\,x^{-1+\epsilon}(1-x)^{-\epsilon}\int_0^1 du\,
u^{-\epsilon}(1-u)^{1-\epsilon}\int_0^1 dy\,y^{-\epsilon}\nonumber\\
&&\cdot\int_0 d\xi\,\delta(\frac{k^2}{x}(1-u)(1-y)+\xi)\cdot\frac{\Gamma(1-\epsilon)}{\sqrt{\pi}\Gamma(\frac{1}{2}-\epsilon)}\int_0^{\pi}d\theta
(\sin\theta)^{-2\epsilon}.
\end{eqnarray}
Finally the integral over $\theta$ can be transformed to an integral from 0 to
1 by substituting $w=\frac{1}{2}(1+\cos{\theta})$, leading to 
\begin{eqnarray}
PS^{ghost}&=&F_{\epsilon}\,|k^2|^{2-2\epsilon}\,x^{-1+\epsilon}(1-x)^{-\epsilon}\int_0^1 du\,
u^{-\epsilon}(1-u)^{1-\epsilon}\int_0^1 dy\,y^{-\epsilon}\nonumber\\
&&\frac{1}{B(\frac{1}{2}-\epsilon,\frac{1}{2}-\epsilon)}\int_0^1
dw\,[w(1-w)]^{-\frac{1}{2}-\epsilon}.
\end{eqnarray}

\end{document}